\documentclass[%
 reprint,
superscriptaddress,
 amsmath,amssymb,
 aps,
pra,
]{revtex4-2}

\usepackage{graphicx}
\usepackage{dcolumn}
\usepackage{bm}
\usepackage{xcolor}
\usepackage{braket}
\usepackage{float}
\usepackage{lipsum}
\usepackage{caption}
\usepackage{subcaption}
\usepackage{float} 

\begin{document}
\preprint{APS/123-QED}

\title{Work extraction from a quantum battery charged through an array of coupled cavities }

\author{I. Beder}
\affiliation{Instituto de F\'isica, Universidade Federal de Alagoas, Macei\'o, Brasil}
\email{Igor.Ribeiro@fis.ufal.br}
\author{D. Ferraro}
\affiliation{Dipartimento di Fisica, Universit\`a di Genova, Via Dodecaneso 33, 16146, Genova, Italy}
\affiliation{SPIN-CNR, Via Dodecaneso 33, 16146, Genova, Italy
}
\email{Dario.Ferraro@unige.it}
\author{P. A. Brand\~ao}
\email{Paulo.Brandao@fis.ufal.br}
\affiliation{Instituto de F\'isica, Universidade Federal de Alagoas, Macei\'o, Brasil}





\date{\today}

\begin{abstract}
We investigate the problem of work extraction from a cavity-based quantum battery that is remotely charged via a transmission line composed of an array of coupled single-mode cavities. For uniform coupling along the line, we show that the ergotropy of the battery, evaluated at the point of maximum power transfer, decreases with the length of the charging line and vanishes beyond a critical size. By carefully engineering the initial state of the charger, nonzero ergotropy can still be harvested even beyond this critical length. We further examine scenarios in which the charging line is initialized in an entangled state, as well as configurations with nonuniform, parabolically varying coupling strengths. In the latter case, we demonstrate that high ergotropy values can be restored, highlighting the potential of spatially engineered interactions to enhance quantum battery performance.
\end{abstract}

\maketitle


\section{Introduction}

It has been more than 120 years since Planck's suggestion that radiation might be formed by quantized packets of energy~\cite{planck1978gesetz} and 60 since the characterization of electromagnetic radiation at the quantum level operated by Glauber~\cite{Glauber63}. However, the quantum control and exploitation of individual photons for technological applications is not at all an outdated topic. Indeed, thanks to the advancement in materials science and nanotechnology, integrated photonics is currently a crucial research topic~\cite{pelucchi2022potential}. In particular, with the progressive development of quantum technologies~\cite{Ezratty24}, quantum properties of light, such as superposition and entanglement, are now being extensively used among the others for the realization of secure quantum information protocols~\cite{Portmann22} and flying qubits for quantum computing~\cite{li2024flying}. 

In recent years, the problem of energy storage and manipulation at the quantum level emerged as a very hot topic in the domain of quantum technologies, leading to the emergence of the new concept of quantum batteries (QBs)~\cite{Bhattacharjee21, Quach23, campaioli2024colloquium}. These kind of devices can be considered as local on-demand energy supplier for quantum computers and quantum sensors, with the potentiality to improve their efficiency and increase their complexity~\cite{Chiribella21, Cioni24, Kurman25}. In this context, photons have played a major role since the early stages. Indeed, they have been considered both as chargers in Dicke QB setups~\cite{Ferraro18, Ferraro19, Quach22, Gemme23, Canzio25, Tibben24, Hymas25} or as way to realize the QB itself~\cite{Seah21, Shaghaghi22, Shaghaghi23, Rodriguez23, Downing24, Andolina25}. In the former case they lead to a super-extensive growth of the the average charging power thank to a phenomenology related to the superradiance~\cite{Kirton19}, while the latter is reminiscent of the micromaser physics largely exploited by S. Haroche and coworkers in their experiments~\cite{Haroche13}.

In practice, large-scale energy distribution and charging protocols, such as those we aim to develop in the next decades to boost quantum technologies, are rarely implemented through direct interaction between the charger and the QB. Instead, properly designed transmission lines are typically employed to mediate the energy transport, as in the case of distributing electricity from a nuclear power plant to a city. Also, in this direction, the photons play a major role. Indeed, motivated by what happens for excitations transfer in light-harvesting photosynthetic systems~\cite{Olaya-Castro08, Sahoo11} or for quantum state transfer between superconducting circuits~\cite{Sillanpää07}, it has been recently demonstrated that a photonic environment can lead to a more efficient and faster energy transfer with respect to alternative quantum mediators~\cite{Crescente22, Crescente24}. According to this, it is essential to consider scenarios where the QB is charged and its energy is extracted via transmission lines. 

In this context, we analyze a simple but paradigmatic model in which the charger and the QB are connected through an array of coupled single-mode cavities \cite{walther2006cavity}. We consider various initial conditions for the charger. When a single photon is initially in the charger, the possibility to efficiently transfer its energy to the QB and extract it as useful work is strongly suppressed by increasing the number of elements composing the transmission line. This situation is only slightly improved by considering a superposition state for the charger or an entangled state along the transmission line. Very remarkably, by properly tuning the coupling among the elements of the transmission line chain it is possible to recover an almost perfect energy transmission and extraction in the same spirit of efficient state transfer in the framework of quantum information~\cite{christandl2004perfect}.

The paper is organized as follows. In Section \ref{sec2} we introduce and diagonalize the theoretical model for a chain composed by $N$ photonic cavities where the first plays the role of charger and the last represents the QB. Section \ref{Applications} consider the energy extraction (in terms of ergotropy) from the QB for the relevant cases of: one single photon initially in the charger, the charger initially in a coherent superposition of vacuum and one-photon states, an entangled state for the mediating chain of cavities and finally in presence of a properly engineered nonuniform coupling among the cavities. Section \ref{Conclusions} is devoted to the conclusions, while one Appendix includes the technical details concerning the diagonalization of the model.

\section{Theoretical Model}\label{sec2}

\begin{figure}[H]
    \centering
    \includegraphics[width=0.45\textwidth]{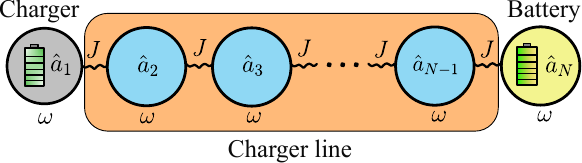}
    \caption{A charger (cavity 1) is connected to a chain of $N-2$ cavities (playing the role of a charging line) having identical coupling constants $J$. The chain is then connected with the QB (cavity $N$). All cavities are assumed to sustain single mode oscillations with frequency $\omega$.}
    \label{system} 
\end{figure}

We consider a system composed of $N$ coupled single-mode cavities, as depicted in Fig. \ref{system}. The first cavity is identified with the charger and the $N$-th cavity is the QB, which will receive the energy from the charger. The $N-2$ remaining cavities form a charging line that mediates the energy transfer between the charger and the QB. The Hamiltonian of the system during this transfer process is given by ($\hbar = 1$)
\begin{equation}\label{hamiltonian}
    \hat{H} = \omega\sum_{p=1}^N\hat{a}_p^{\dagger}\hat{a}_p + J\sum_{p=1}^{N-1}(\hat{a}_p^{\dagger}\hat{a}_{p+1} + \hat{a}_{p+1}^{\dagger}\hat{a}_p),
\end{equation}
where $\omega$ is the frequency of each cavity and $\hat{a}_p$ is the annihilation operator of the $p$-th cavity mode. It is possible to diagonalize the Hamiltonian in Eq.~\eqref{hamiltonian} and find the exact Heisenberg solutions for the operators $\hat{a}_p(t)$ (see Appendix A for the details). They are given by
\begin{equation}\label{apmain}
    \hat{a}_p(t) = \sum_{r=1}^N\sum_{k=1}^N S(r,k)S(r,p)e^{-i\Omega_r t}\hat{a}_k(0),
\end{equation}
where $\Omega_r = \omega + 2J\cos[r\pi/(N+1)]$ and
\begin{equation}\label{smn}
    S(m,n) = \sqrt{\frac{2}{N+1}}\sin\left( \frac{mn\pi}{N+1} \right).
\end{equation}

To find an expression for the state ket $\ket{\psi(t)}$ at time $t$, let us assume that the initial state of the system is given by $\ket{\psi(0)} = \ket{\phi_1}\ket{0_2}\ket{0_3}\ldots\ket{0_N}$, where the quantum state of the charger (cavity 1) is $\ket{\phi_1}$ and the other cavities, including the QB $\ket{0_N}$, are initially empty (in their vacuum states). Then, if $\hat{U}(t) = e^{-i\hat{H}t}$ is the time evolution operator,
\begin{equation}\label{psit}
    \begin{split}
        \ket{\psi(t)} &= \hat{U}(t) \ket{\psi(0)} \\
        &= \sum_{n_1 = 0}^{\infty}\frac{c_{n_1}}{(n_1!)^{1/2}}\hat{U}(t)(\hat{a}_1^{\dagger})^{n_1}\ket{\mathbf{0}}\\
        &= \sum_{n_1 = 0}^{\infty}\frac{c_{n_1}}{(n_1!)^{1/2}} [\hat{a}_1^{\dagger}(-t)]^{n_1}\ket{\mathbf{0}},
    \end{split}
\end{equation}
where the operator $\hat{a}_1$ in the second line of Eq. \eqref{psit} is expressed in the Schr\"odinger picture, $\ket{\mathbf{0}} = \ket{0_1}\ket{0_2}\ldots\ket{0_N}$ and $\{ c_{n_1} \}_{n_1 = 0}^{\infty}$ are the expansion coefficients of $\ket{\phi_1}$ in the number basis of the charger. The passage from the second to the third line in Eq. \eqref{psit} was accomplished by using the evolution of the creation operator in Heisenberg representation $\hat{a}_1^{\dagger}(-t) = \hat{U}(t)\hat{a}_1^{\dagger}\hat{U}^{\dagger}(t)$ and the fact that $\hat{U}(t)\ket{\mathbf{0}} = \ket{\mathbf{0}}$. 

\section{Applications} \label{Applications}

This Section is devoted to the applications of the general formalism developed in Sec. \ref{sec2} by focusing on specific initial states of the charger and investigating the energy, power and ergotropy of the QB (see below for the proper definitions of these figures of merit). 

We are interested in the amount of energy that can be stored into the QB and the part of this energy which can be extracted as useful work through unitary operations (know as ergotropy \cite{allahverdyan2004maximal,alicki2013entanglement,touil2021ergotropy}) at the end of the transfer protocol. The charging protocol is the following: Initially, the charger and the QB are disconnected from the array of coupled cavities and remain in the states $\ket{\phi_1}$ and $\ket{0_N}$, respectively. At time $t = 0$, they are connected through the array of coupled cavities, as depicted in Fig. \ref{system}, and interact until $t = \tau$. The battery is then disconnected from the array and we determine the energy $E_N(\tau)$ stored into the battery and ergotropy $\mathcal{E}_N(\tau)$ that remains after the interaction.

Subsection \ref{single} considers the case where the charger initially holds a single photon. Subsection \ref{superposition} deals with an initial superposition state of the charger, Section \ref{entangled} analyzes the case where the transmission line is initially in an entangled $W$ state and Section \ref{nonuniform} is devoted to an example involving nonuniform couplings.

\subsection{Charger having a single photon}\label{single}

Let us start with the simplest case in which the charger initially holds one photon, $\ket{\psi(0)} = \ket{1_1}\ket{0_2,...,0_{N-1}}\ket{0_N}$, that is $c_{n_1} = \delta_{n_1,1}$, where $\delta_{m,n}$ is the Kronecker delta. In this case, the state of the system at time $t$ can be obtained from Eq. \eqref{psit},
\begin{equation}\label{psi1foton}
    \ket{\psi(t)} = \sum_{k=1}^N G_k(t)
    \ket{\mathbf{1}_k},
\end{equation}
where 
\begin{equation}
    G_k(t) = \sum_{r=1}^NS(r,k)S(r,1)e^{-i\Omega_r t}
\end{equation}
and $\ket{\mathbf{1}_k} = \ket{0_1}\ldots\ket{1_k}\ldots\ket{0_N}$ is the state where the $k$-th cavity has a single photon excitation. Thus, even though the initial state is separable, the interaction between the cavities generates an entangled state of the whole system.

From Eq. \eqref{psi1foton} the density operator $\hat{\rho}(t) = \ket{\psi(t)}\bra{\psi(t)}$ can be calculated,
\begin{equation}
    \hat{\rho}(t) = \sum_{kk'}G_k(t)G^*_{k'}(t)\ket{\mathbf{1}_k}\bra{\mathbf{1}_{k'}},
\end{equation}
and the density operator $\hat{\rho}_N(t)$ for the battery is then obtained by tracing out the states of the charger and of the other cavities (from $1$ to $N-1$),
\begin{equation}\label{rho1t}
    \begin{split}
        &\hat{\rho}_N(t) = \text{Tr}_{1,...,N-1}\Big[\hat{\rho}(t)\Big] \\
        &= \Big[1-|G_N(t)|^2\Big] \ket{0_N}\bra{0_N} + |G_N(t)|^2\ket{1_N}\bra{1_N}.
        \end{split}
\end{equation}
The expression in Eq.~\eqref{rho1t} shows that the state of the battery at any time has no coherences between the states $\ket{0_N}$ and $\ket{1_N}$. A situation with nonzero coherences will be considered in the next section. 

\begin{figure}[H]
\includegraphics[width=0.9\linewidth]{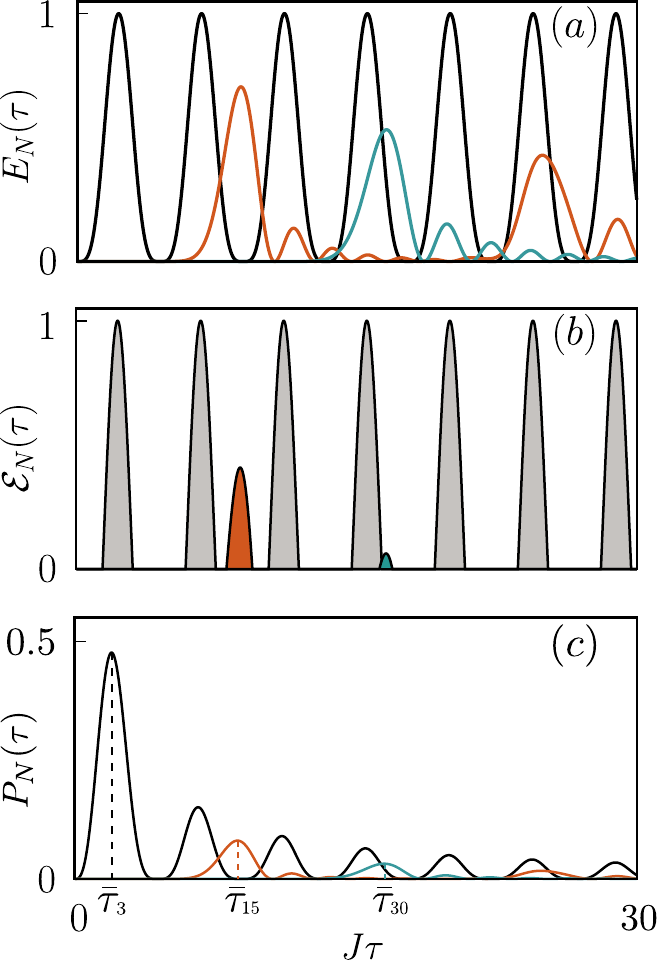}
    \centering
\caption{(a) Energy $E_N(\tau)$, (b) ergotropy $\mathcal{E}_N(\tau)$ and (c) averaged charging power $P_N(\tau)$ as a function of $J\tau$ for (black) $N = 3$, (orange) $N = 15$ and (cyan) $N = 30$ cavities. The ratio $\omega/J = 1$ is considered for all the simulations performed in this paper. }
\label{fig2}
\end{figure}

At the end of the charging protocol, $t = \tau$, the energy of the battery, given by $E_N(\tau) = \text{Tr}[\hat{H}_N\hat{\rho}_N(\tau)]$ where $\hat{H}_N = \omega \hat{a}_N^{\dagger}\hat{a}_N$ is the Hamiltonian of the battery~\cite{andolina2018charger, Cavaliere25}, can be calculated explicitly,
\begin{equation}\label{energysingle}
    \begin{split}
        E_N(&\tau) =\omega|G_N(\tau)|^2 \\
        &= \omega\Bigg| \sum_{r=1}^NS(r,N)S(r,1)e^{-i\Omega_r t}  \Bigg|^2. 
    \end{split}
\end{equation}
Figure \ref{fig2}(a) shows $E_N(\tau)$ as a function of $Jt$ for $N = 3$ (black line), $N = 15$ (orange line) and $N = 30$ (cyan line). Notice that, hereafter, all quantities are plotted in the proper units of the angular frequency $\omega$. One sees that, if a single cavity mediates the charging of the battery ($N = 3$), the energy oscillates periodically and the battery can fully absorb the initial energy of the charger. For $N = 15$ and $N = 30$, the periodicity of $E_N(\tau)$ is lost and the oscillations of the energy stored in the battery are progressively suppressed. This is expected because the (finite) initial energy present in the charger becomes distributed throughout many cavities, each sharing a relatively small amount of it.

Not all of the energy $E_N(\tau)$ can be transformed into work. There are states of the battery, called passive states and indicated in the following as $\hat{\sigma}_N$, that do not allow energy extraction. A global measure of the amount of useful energy contained in the battery is given by the ergotropy~\cite{allahverdyan2004maximal},
\begin{equation}
    \mathcal{E}(\tau) = \text{Tr}[\hat{H}_B\hat{\rho}_N(\tau)] - \text{Tr}[\hat{H}_B\hat{\sigma}_N(\tau)].
\end{equation}
The quantity $\mathcal{E}$ represents the maximum amount of energy that can be transformed into work by applying unitary cyclic transformations to $\hat{\rho}_N(\tau)$ \cite{allahverdyan2004maximal,alicki2013entanglement,touil2021ergotropy, Morone23, Castellano24, Elyasi25}. The passive states $\hat{\sigma}_N(\tau)$ associated to it, can be constructed from the spectral decomposition of the density operator, $\hat{\rho}_N = \sum_{k=0}^M\lambda_k\ket{\lambda_k}\bra{\lambda_k}$ ($\lambda_0 > \lambda_1 > ... > \lambda_M$),  and the spectral decomposition of the Hamiltonian $\hat{H}_N = \sum_{k = 0}^L \varepsilon_k\ket{\varepsilon_k}\bra{\varepsilon_k}$ ($\varepsilon_0 < \varepsilon_1 < ... < \varepsilon_L$), where $\hat{H}_N\ket{\varepsilon_j} = \varepsilon_j\ket{\varepsilon_j}$. It is given by $\hat{\sigma}_N = \sum_{k = 0}^{Q}\lambda_k\ket{\varepsilon_k}\bra{\varepsilon_k}$. No work can be extracted from this state because the ordering of $\lambda_k$ assigns the highest population $\lambda_0$ to the lowest energetic state $\ket{\varepsilon_0}$. In this context, thermal states are one example of passive states (but not all passive states are thermal \cite{campaioli2024colloquium}).

Returning to our system, it is straightforward to demonstrate that the ergotropy is given by $\mathcal{E}(\tau) = \omega [ 2|G_N(\tau)|^2 - 1 ]\Theta [ |G_N(\tau)|^2 - \frac{1}{2} ]$, where $\Theta(a)$ is equal to one if $a>0$ and zero if $a<0$. Figure \ref{fig2}(b) shows $\mathcal{E}(\tau)$ as a function of $J\tau$ for the same values of $N$ considered for the energy. Also in this case, one sees that the ergotropy is a periodic function for $N = 3$ and that it decreases drastically as $N$ increases. 

To compare charging performances at different number $N$ of cavities, we follow previous approaches by focusing on the interaction time $\tau = \bar{\tau}$ for which the delivered averaged charging power $P_N(\tau) = E_N(\tau)/\tau$ is maximum \cite{Carrega16, Andolina19}. Notice that $\bar{\tau}$ depends on the number of cavities present in the charging line. Figure \ref{fig2}(c) shows $P_N(\tau)$ for $N = 3$, $N = 15$ and $N = 30$ cavities. 

\begin{figure}[H]
\includegraphics[width=0.9\linewidth]{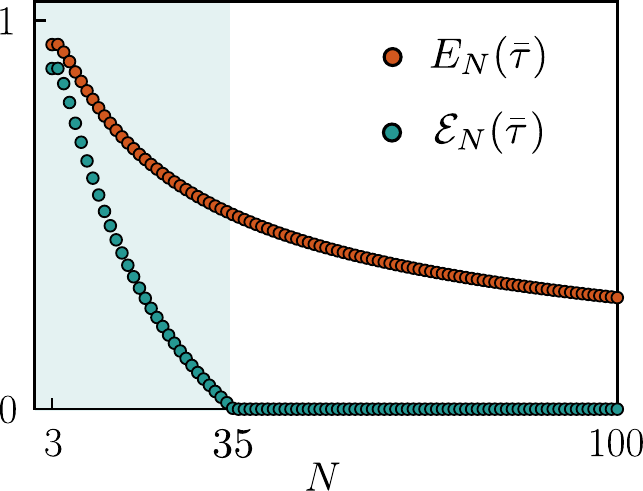}
    \centering
\caption{Energy $E_N(\bar{\tau})$ (orange dots) and ergotropy $\mathcal{E}_N(\bar{\tau})$ (cyan dots), both evaluated at the time where the maximum aveeraged charging power is transferred $\bar{\tau}$, as a function of $N$. For this set of parameters the ergotropy vanishes for $N \geq 35$.}
\label{fig3}
\end{figure}

Figure \ref{fig3} displays $E_N(\bar{\tau})$ and $\mathcal{E}_N(\bar{\tau})$ as a function of $N$. Despite the nonzero values for the energy $[E_N(\bar{\tau})]$ for all $N$ considered, we numerically observe that the ergotropy vanishes if $N \geq 35$. This means that no useful energy can be extracted from the QB if the number of cavities mediating the charging protocol is larger than the critical number $N_c = 35$. Since the number of cavities is directly related to the length of the charging line, the charging of a QB very distant from the charger can be severely restricted when a single photon is initially in the latter.

\subsection{Charger initially in a superposition state}\label{superposition}

In the previous part, we have demonstrated a serious limitation for the ergotropy of a QB charged by a transmission line. Here and in the next subsection, we will explore the stored energy and the ergotropy of the QB by judiciously changing the initial condition of the system. Let us start by considering a superposition state for the number of photons initially present in the charger, namely
\begin{equation}
\label{psi0super}
    \ket{\phi_1} = \frac{1}{\sqrt{1 + \beta^2}}\Big( \ket{1_1} + \beta e^{i\varphi}\ket{0_1}\Big),
\end{equation}
where $\beta \geq 0$ and $\varphi \in [0,2\pi]$. The expansion coefficients $c_{n_1}$ in the number basis are then given by $c_0 = \beta e^{i\varphi}/\sqrt{1+\beta^2}$, $c_1 = 1/\sqrt{1+\beta^2}$ and $c_{n_1} = 0$ for $n_1 > 1$. At first glance, the initial condition in Eq.~\eqref{psi0super} seems a bad choice to increase the ergotropy since the charger has now a nonzero probability of being found in the fundamental energy state $\ket{0_1}$ which cannot contribute to the charging of the QB. However, we will demonstrate that this is indeed not the case. Eq.~\eqref{psit} is now used to obtain the state of the system at a given time $t$: $\ket{\psi(t)} = c_0\ket{\mathbf{0}} + c_1\sum_{k=1}^N G_k(t)\ket{\{ 1 \}_k}$. It is straightforward to obtain the density operator for the QB,
\begin{equation}\label{rhoent}
\hat{\rho}_N(t) = 
    \begin{bmatrix}
        |c_0|^2 + c_1^2[1 - |G_N(t)|^2] & c_0c_1 G_N^*(t) \\
        c_0^*c_1 G_N(t) & c_1^2|G_N(t)|^2
    \end{bmatrix},
\end{equation}
which satisfies Tr$[\hat{\rho}_N(t)]$ = 1 as expected. The matrix is also positive and Hermitian (note that $c_1$ is real-valued). Differently from the previous case, here the state in Eq.~\eqref{rhoent} has nonzero off-diagonal elements (coherences).

The construction of the passive state $\hat{\sigma}_N$ proceeds in exactly the same way as before. The eigenvalues of Eq.~\eqref{rhoent} are 
\begin{equation}
    \lambda_{0,1} = \frac{1}{2}\Bigg( 1 \pm \sqrt{1 - 4c_1^4|G_N|^2(1 - |G_N|^2)} \Bigg),
\end{equation}
where the plus (minus) sign is associated with $\lambda_0$ ($\lambda_1$). The passive state is then given by $\hat{\sigma}_N = \lambda_0\ket{\varepsilon_0}\bra{\varepsilon_0}+ \lambda_1\ket{\varepsilon_1}\bra{\varepsilon_1}$ and the ergotropy is given by 
\begin{equation}\label{maxesuper}
\begin{split}
    \mathcal{E}&(\tau) = \omega c_1^2|G_N(\tau)|^2  \\
    &+ \frac{\omega}{2}\Bigg[ \sqrt{1-4c_1^4|G_N(\tau)|^2(1-|G_N(\tau)|^2)} - 1 \Bigg] .
\end{split}
\end{equation}
Notice that the energy of the battery is given by $E_N(\tau) = \omega c_1^2|G_N(\tau)|^2$ and, since max($c_1$) = 1, the energy $E_N(\tau)$ for $c_1 < 1$ is always smaller than the one evaluated in the previous section ($c_1 = 1)$. 

\begin{figure}[H]
\includegraphics[width=\linewidth]{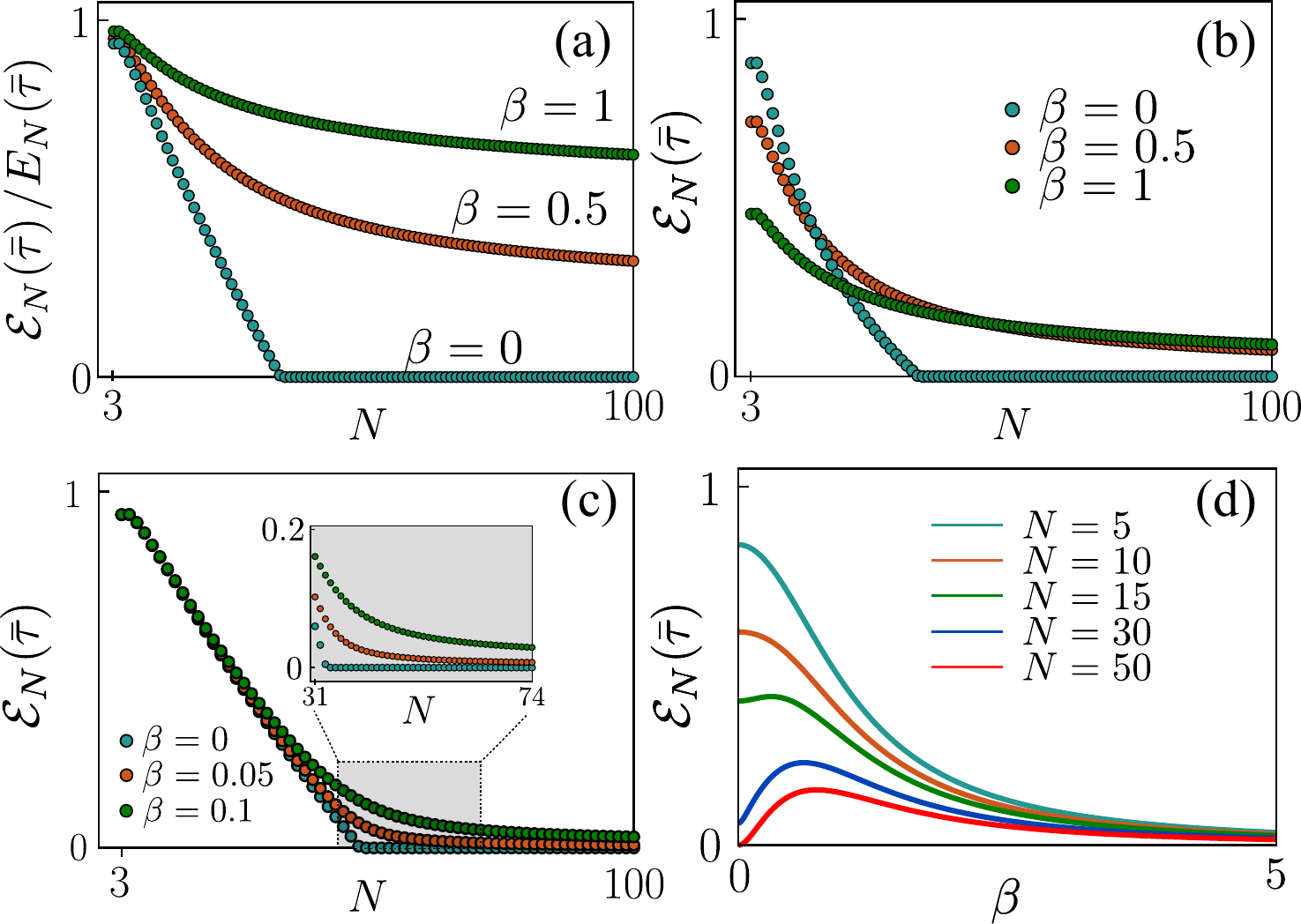}
    \centering
\caption{(a) Ratio [$\mathcal{E}_N(\bar{\tau})/E_N(\bar{\tau})$] as a function of $N$ for three values of $\beta$. (b) Plot of the ergotropy $\mathcal{E}(\bar{\tau})$ as a function of $N$ for the same values of $\beta$ shown in part (a). (c) Same as in part (b) but with small values of $\beta$, indicating that as soon as $\beta \neq 0$, the ergotropy does not drop to zero at any $N$ and remains finite, even if very small. (d) Ergotropy as a function of $\beta$ for different values of $N$.}
\label{fig4}
\end{figure}

Figure \ref{fig4}(a) shows the ratio $\mathcal{E}_N(\bar{\tau})/E_N(\bar{\tau})$ as a function of $N$ for $\beta = 0$, $0.5$ and $1$. The line with $\beta = 0$ (cyan) corresponds to what was derived in the previous section, where the ergotropy vanishes for $N \geq 35$. However, the ratio $\mathcal{E}_N(\bar{\tau})/E_N(\bar{\tau})$ increases for nonzero $\beta$. This suggests that coherences in the state of the charger can help to increase the fraction of useful energy present in the QB. Unfortunately, as discussed in the last paragraph, it happens at cost of the energy available. Figure \ref{fig4}(b) shows the behavior of the ergotropy as a function of $N$ for different values of $\beta$ (to be compared with Figure \ref{fig3}).

Figure \ref{fig4}(c) reports the dependence of the ergotropy as a function of $N$ as the parameter $\beta$ increases. We found that there is an abrupt change in the behavior in the sense that as soon as $\beta \neq 0$, the ergotropy never vanishes for any $N$. Figure \ref{fig4}(d) displays the dependence of the ergotropy on the parameter $\beta$. Quite remarkably, for some values of $N$, the ergotropy $\mathcal{E}_N(\bar{\tau})$ displays a maximum for $\beta \neq 0$. For example, in the case $N = 50$ the ergotropy vanishes for $\beta = 0$ (see Fig. \ref{fig3}) but it has a maximum for $\beta 
 \neq 0$ (see red curve).  We thus find that there is a tradeoff between the cases where $\beta = 0$ and $\beta \neq 0$ in the sense that the ergotropy can be made nonzero for $N \geq 35$ but only at the expense of decreasing the overall energy in the battery ($\beta \neq 0$). 

\begin{figure}[H]
\includegraphics[width=\linewidth]{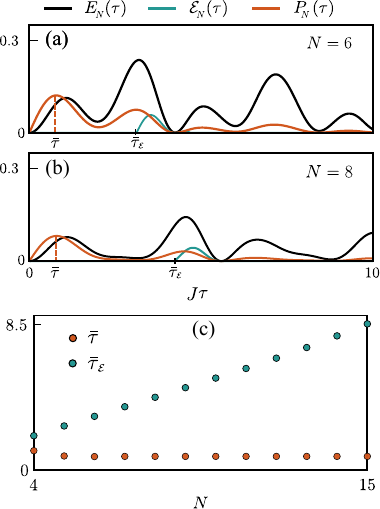}
    \centering
\caption{Energy $E_N(\tau)$, ergotropy $\mathcal{E}_N(\tau)$ and averaged charging power $P_N(\tau)$ as a function of $J\tau$ for (a) $N = 6$ and (b) $N=8$ cavities. (c) Times $\bar{\tau}$ and $\bar{\tau}_{\mathcal{E}}$ at which the first maximum value of power and ergotropy occur, respectively. 
The charging line is initially in the so called $W$ state given by  $\ket{\Phi_{\text{array}}} = \sum_{r=2}^{N-1}\ket{\mathbf{1}_r}/\sqrt{N-2}$.}
\label{fig5}
\end{figure}

\subsection{A chain of entangled cavities}\label{entangled}

Our final discussion involving uniform couplings addresses the case where the array of coupled cavities mediating the energy transfer is initially in an entangled state. Recently, it has been shown that coupled arrays of entangled cavities can enhance the charging efficiency of the QB \cite{ma2024enhancing}. Let us assume then the initial state $\ket{\psi(0)} = \ket{1_1}\ket{\Phi_{\text{array}}}\ket{0_N}$, where $\ket{\Phi_{\text{array}}} = \sum_{r=2}^{N-1}d_r\ket{\bm{1}_r }$, with $\sum_{r=2}^{N-1}|d_r|^2 = 1$, is the state of the charging line. The time-evolved state of the system is now given by 
\begin{equation}
    \ket{\psi(t)} = \sum_{r=2}^{N-1}d_r\hat{a}_1^{\dagger}(-t)\hat{a}_r^{\dagger}(-t)\ket{\mathbf{0}}.
\end{equation}
Using Eq. \eqref{apmain} and the definition
\begin{equation}
\begin{split}
    \mathcal{G}_{p,q}(t) &= \sum_{r=2}^{N-1}d_r\sum_{l,m=1}^{N} S(l,p)S(l,1)S(m,q)S(m,r) \\
    & \times e^{-i(\Omega_l + \Omega_m)t},
\end{split}
\end{equation}
the state $\ket{\psi(t)}$ can be written in terms of the double excitation subspace basis,
\begin{equation}
    \begin{split}
    &\ket{\psi(t)} = \sum_{p,q=1}^{N} \mathcal{G}_{p,q}(t)\ket{\bm{1}_q,\bm{1}_p} \\
    &= \sum_{p\neq q}^{N} \mathcal{G}_{p,q}(t)\ket{\bm{1}_q,\bm{1}_p} + \sum_{r=1}^{N} \mathcal{G}_{r,r}(t)\ket{\bm{2}_r}
    \end{split}
\end{equation}
where $\ket{\bm{1}_q,\bm{1}_p}$ means that one photon is located at cavity $q$ and the other at cavity $p\neq q$ and $\ket{\bm{2}_q}$ characterizes two photons at the same cavity $q$. The density operator for the battery at $t = \tau$ is now given by
\begin{equation}
    \begin{split}
        \rho_N(\tau) &= \Bigg[ 1 - \sum_{n=1}^{N}|\mathcal{G}_{n,N}(\tau)|^2 \Bigg]\ket{0_N}\bra{0_N} \\
        &+ \sum_{n=1}^{N-1}|\mathcal{G}_{n,N}(\tau)|^2\ket{1_N}\bra{1_N}\\
        &+ |\mathcal{G}_{N,N}(\tau)|^2\ket{2_N}\bra{2_N}.
    \end{split}
\end{equation}
from which the energy, ergotropy and the power can be extracted.

Figure \ref{fig5} plots $E_N$, $\mathcal{E}_N$ and $P_N$ as a function of $J\tau$ for $N = 6$, $N = 8$ and $N = 10$ cavities. We consider the coefficients $d_r = 1/\sqrt{N-2}$ so that $\ket{\Phi_{\text{array}}}$ becomes the well-known $W$ state \cite{dur2000three}. It is evident from these results that the time at which maximum power is achieved never coincides with a nonzero value for the ergotropy. Therefore, the ergotropy $\mathcal{E}_N(\bar{\tau})$ (at maximum power transfer) is always zero if the charging line is initially in the entangled $W$ state. 

\subsection{Nonuniform couplings}\label{nonuniform}

So far, irrespective of the initial state of the charger or the charging line, the overall energy and ergotropy of the battery decreases as the number of cavities mediating the charging process increases. As already mentioned, this behavior is expected because the electromagnetic energy has a tendency to spread throughout the cavities. This puts a limit in the applicability of the discussed charging line protocol. However, one may ask if there are nonuniform configurations of the coupling coefficients such that the charging performance can be improved. 

In this direction, we consider a charging model in which the coupling coefficients are given by 
\begin{equation}\label{couplingparabolic}
    J_p = J\sqrt{p(N-p)} \quad (p=1,...,N-1),
\end{equation}
where $J$ is a constant parameter. This model has been used to demonstrate perfect quantum state transfer between distinct spins in spin chains \cite{christandl2004perfect}. In this case, the Hamiltonian is still given by Eq. \eqref{hamiltonian} except that $J \rightarrow J_p = J\sqrt{p(N-p)}$ and must be put inside the summation over $p$ in the interaction term. In the Heisenberg picture, the operators $a_p(t)$ are now written as $\hat{a}_p(t) = \sum_{l=1}^N A_{pl}(t)\hat{a}_l(0)$,
where the scalar functions $A_{pl}(t)$ satisfy the set of coupled differential equations
\begin{equation}\label{systemapl}
    \begin{split}
        i\frac{dA_{1l}}{dt} & = J_1 A_{2l}, \\
        i\frac{dA_{pl}}{dt} &= J_pA_{p+1,l} + J_{p-1}A_{p-1,l}\quad (p \neq 1,N) , \\
        i\frac{dA_{Nl}}{dt} &= J_{N-1}A_{N-1,l}.
    \end{split}
\end{equation}
After numerically solving the system in Eq.\eqref{systemapl}, the construction of the state $\ket{\psi(t)}$ is then obtained by using again Eq. \eqref{psit}.

\begin{figure}[H]
\includegraphics[width=\linewidth]{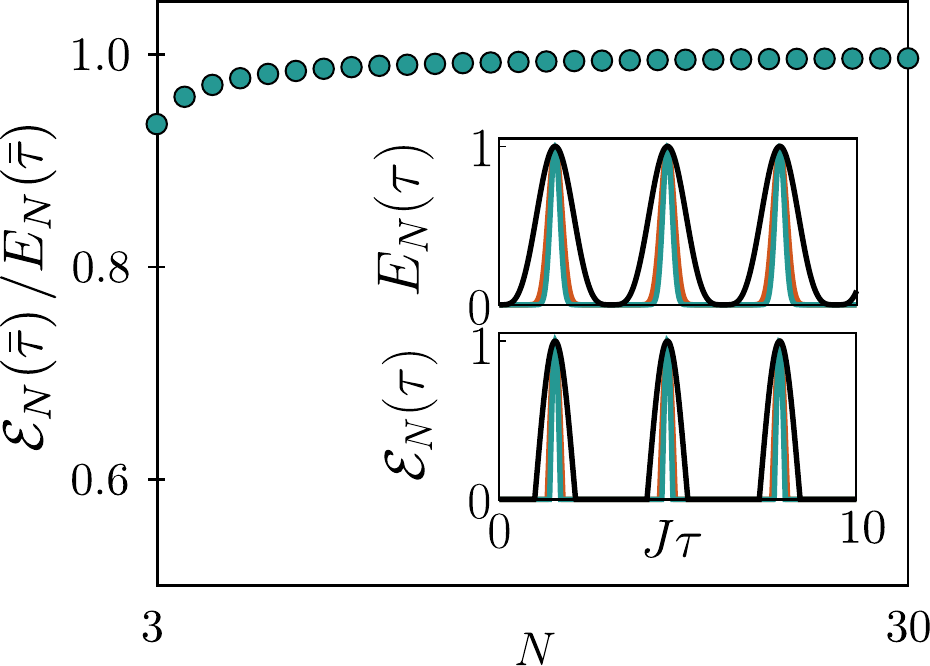}
    \centering
\caption{Ratio $\mathcal{E}_N(\bar{\tau})/E_N(\bar{\tau})$, evaluated at the time $\bar{\tau}$ at which the maximum power is achieved, as a function of $N$. The inset shows the energy $E_N(\tau)$ and ergotropy $\mathcal{E}_N(\tau)$ as a function of $J\tau$, for (black) $N = 3$, (orange) $N = 15$ and (cyan) $N = 30$ cavities.} 
\label{figparabola}
\end{figure}

Let us then reconsider the initial situation where the charger holds a single photon excitation $\ket{1_1}$. In this case, the density operator for the QB is 
\begin{multline}
    \hat{\rho}_N(t) = \Big[ 1 - |A_{1N}(-t)|^2 \Big]\ket{0_N}\bra{0_N}\\
    + |A_{1N}(-t)|^2\ket{1_1}\bra{1_1}.
\end{multline}
Figure \ref{figparabola} shows the ratio $\mathcal{E}_N(\bar{\tau})/E_N(\bar{\tau})$ as a function of $N$. Remarkably, this ratio increases as $N$ increases. The insets show the energy and ergotropy as a function of $J\tau$ for $N = 3$, $N = 15$ and $N = 30$ cavities. The slight decrease of the ratio for small values of $N$ occur because the time $\bar{\tau}$ at maximum power transfer does not exactly match the maximum ergotropy. But they asymptotically coincide in the $N\rightarrow \infty$ limit. Notice that the observed saturation of the ratio occur at expensive of the time scale for which energy and ergotropy are significantly different from zero, that shrinks as $N$ increases.

To conclude, we want to point out that the proposed scheme to charge quantum batteries via next-neighbor couplings can be experimentally realized with the current technology. In particular, a linear chain of coupled superconducting qubits with the coupling model introduced in Eq.~\eqref{couplingparabolic} has been devised to address the perfect quantum state transfer protocol \cite{PhysRevApplied.10.054009}. In this case, an ac magnetic flux is used to parametrically modulate the frequency of each superconducting qubit, which creates an effective non-uniform coupling strength. The same coupling model can be experimentally engineered using the polarization states of light as qubits, and an array of single-mode coupled waveguides \cite{chapman2016experimental}.

\section{Conclusions} \label{Conclusions}

We have explored the charging of a quantum battery, defined as the quantized electromagnetic field inside a single-mode cavity, by a charger that is located far away from the battery. The energy transfer among them is not direct but mediated by a chain of cavities. We have considered  various possible initial conditions for the charger. In particular, when a single photon is initially present in the charger, the possibility to efficiently transfer its energy to the QB and extract it as useful work progressively decreases as the number of elements composing the transmission line increases. This situation is partially improved by considering a superposition state for the charger or an entangled state for the cavities composing the transmission line. Conversely, a saturation of the efficiency of the device concerning the energy extraction can be achieved by properly tuning the coupling among the elements of the transmission line chain. This could lead to interesting application of the presented scheme also in the broader framework of spin-chain~\cite{Le18, Zhao21, Grazi24, Catalano24} or topological~\cite{Lu25} quantum batteries. 
 
Note added: While finalizing this manuscript, we became aware of a related work reporting a similar idea in the framework of spin-chain quantum batteries \cite{murphy2025ergotopy}. In their work, the authors placed greater emphasis on the role of disorder in non-uniform couplings, while not addressing the transfer of ergotropy in comparison with the time of maximum power, as we did. Despite differences in the approach, our findings are in agreement with their analyses.

\begin{acknowledgments}
P. A. B. acknowledges the financial support of CNPq (Conselho Nacional de Desenvolvimento Cient\'ifico e Tecnol\'ogico), I. B. acknowledges the financial support of CAPES (Coordenação de Aperfeiçoamento de Pessoal de Nível Superior) and D.F. acknowledges the contribution of the European Union-NextGenerationEU through the “Quantum Busses for Coherent Energy Transfer” (QU
BERT) project, in the framework of the Curiosity Driven
 2021 initiative of the University of Genova and the support from the project PRIN 2022
2022XK5CPX (PE3) SoS-QuBa- “Solid State Quantum
 Batteries: Characterization and Optimization” funded
 within the programme “PNRR Missione 4- Componente
 2- Investimento 1.1 Fondo per il Programma Nazionale
 di Ricerca e Progetti di Rilevante Interesse Nazionale
 (PRIN)”, funded by the European Union- Next Gen
eration EU.
\end{acknowledgments}

\section{Appendix A}

The Hamiltonian in Eq.~\eqref{hamiltonian} can be diagonalized in terms of collective operators $\hat{c}_q$, related to $\hat{a}_q$ by
\begin{equation}\label{apend1}
    \hat{a}_p = \sum_{q=1}^N S(q,p)\hat{c}_q,
\end{equation}
where $S(q,p)$ are real-valued parameters. In terms of $\hat{c}_q$, the Hamiltonian is written as
\begin{equation}
    \begin{split}
        &\hat{H} = \omega\sum_{p,q,r=1}^NS(q,p)S(r,p)\hat{c}_q^{\dagger}\hat{c}_r \\
        &+ J\sum_{q,r=1}^N \hat{c}_q^{\dagger}\hat{c}_r \sum_{p=1}^{N-1}\Big[ S(q,p)S(r,p+1)\\
        &+ S(r,p)S(q,p+1) \Big]
    \end{split}.
\end{equation}
If $S(m,n)$ is defined according to Eq. \eqref{smn}, the identities
\begin{equation}
\begin{split}
    \sum_{p=1}^{N-1}\Big[ &S(q,p)S(r,p+1) + S(r,p)S(q,p+1) \Big] \\
    &= 2\delta_{q,r}\cos\Bigg(\frac{r\pi}{N+1} \Bigg)
    \end{split}
\end{equation}
and
\begin{equation}
    \sum_{p=1}^N S(p,q)S(p,r) = \delta_{q,r}
\end{equation}
can be used to write the Hamiltonian in the diagonal form
\begin{equation}
    \hat{H} = \sum_{r = 1}^N \Omega_r \hat{c}_r^{\dagger}\hat{c}_r,
\end{equation}
where $\Omega_r = \omega + 2J\cos[r\pi/(N+1)]$. Thus, the system is equivalent to $N$ \textit{uncoupled} harmonic oscillators having distinct frequencies $\Omega_r$. In the Heisenberg picture, the solution for $\hat{c}_k(t)$ is  obtained directly from the Heisenberg equations,
\begin{equation}\label{apend2}
    \hat{c}_k(t) = \hat{c}_k(0)e^{-i\Omega_k t}.
\end{equation}
Expression \eqref{apend2} can now be substituted back into Eq. \eqref{apend1}. The resulting expression is given by Eq. \eqref{apmain} in the main text.

\appendix


\bibliography{apssamp}

\begin{thebibliography}{54}%
\makeatletter
\providecommand \@ifxundefined [1]{%
 \@ifx{#1\undefined}
}%
\providecommand \@ifnum [1]{%
 \ifnum #1\expandafter \@firstoftwo
 \else \expandafter \@secondoftwo
 \fi
}%
\providecommand \@ifx [1]{%
 \ifx #1\expandafter \@firstoftwo
 \else \expandafter \@secondoftwo
 \fi
}%
\providecommand \natexlab [1]{#1}%
\providecommand \enquote  [1]{``#1''}%
\providecommand \bibnamefont  [1]{#1}%
\providecommand \bibfnamefont [1]{#1}%
\providecommand \citenamefont [1]{#1}%
\providecommand \href@noop [0]{\@secondoftwo}%
\providecommand \href [0]{\begingroup \@sanitize@url \@href}%
\providecommand \@href[1]{\@@startlink{#1}\@@href}%
\providecommand \@@href[1]{\endgroup#1\@@endlink}%
\providecommand \@sanitize@url [0]{\catcode `\\12\catcode `\$12\catcode `\&12\catcode `\#12\catcode `\^12\catcode `\_12\catcode `\%12\relax}%
\providecommand \@@startlink[1]{}%
\providecommand \@@endlink[0]{}%
\providecommand \url  [0]{\begingroup\@sanitize@url \@url }%
\providecommand \@url [1]{\endgroup\@href {#1}{\urlprefix }}%
\providecommand \urlprefix  [0]{URL }%
\providecommand \Eprint [0]{\href }%
\providecommand \doibase [0]{https://doi.org/}%
\providecommand \selectlanguage [0]{\@gobble}%
\providecommand \bibinfo  [0]{\@secondoftwo}%
\providecommand \bibfield  [0]{\@secondoftwo}%
\providecommand \translation [1]{[#1]}%
\providecommand \BibitemOpen [0]{}%
\providecommand \bibitemStop [0]{}%
\providecommand \bibitemNoStop [0]{.\EOS\space}%
\providecommand \EOS [0]{\spacefactor3000\relax}%
\providecommand \BibitemShut  [1]{\csname bibitem#1\endcsname}%
\let\auto@bib@innerbib\@empty
\bibitem [{\citenamefont {Planck}(1978)}]{planck1978gesetz}%
  \BibitemOpen
  \bibfield  {author} {\bibinfo {author} {\bibfnamefont {M.}~\bibnamefont {Planck}},\ }\href@noop {} {\emph {\bibinfo {title} {{\"U}ber das gesetz der energieverteilung im normalspektrum}}}\ (\bibinfo  {publisher} {Springer},\ \bibinfo {year} {1978})\BibitemShut {NoStop}%
\bibitem [{\citenamefont {Glauber}(1963)}]{Glauber63}%
  \BibitemOpen
  \bibfield  {author} {\bibinfo {author} {\bibfnamefont {R.~J.}\ \bibnamefont {Glauber}},\ }\bibfield  {title} {\bibinfo {title} {The quantum theory of optical coherence},\ }\href {https://doi.org/10.1103/PhysRev.130.2529} {\bibfield  {journal} {\bibinfo  {journal} {Phys. Rev.}\ }\textbf {\bibinfo {volume} {130}},\ \bibinfo {pages} {2529} (\bibinfo {year} {1963})}\BibitemShut {NoStop}%
\bibitem [{\citenamefont {Pelucchi}\ \emph {et~al.}(2022)\citenamefont {Pelucchi}, \citenamefont {Fagas}, \citenamefont {Aharonovich}, \citenamefont {Englund}, \citenamefont {Figueroa}, \citenamefont {Gong}, \citenamefont {Hannes}, \citenamefont {Liu}, \citenamefont {Lu}, \citenamefont {Matsuda} \emph {et~al.}}]{pelucchi2022potential}%
  \BibitemOpen
  \bibfield  {author} {\bibinfo {author} {\bibfnamefont {E.}~\bibnamefont {Pelucchi}}, \bibinfo {author} {\bibfnamefont {G.}~\bibnamefont {Fagas}}, \bibinfo {author} {\bibfnamefont {I.}~\bibnamefont {Aharonovich}}, \bibinfo {author} {\bibfnamefont {D.}~\bibnamefont {Englund}}, \bibinfo {author} {\bibfnamefont {E.}~\bibnamefont {Figueroa}}, \bibinfo {author} {\bibfnamefont {Q.}~\bibnamefont {Gong}}, \bibinfo {author} {\bibfnamefont {H.}~\bibnamefont {Hannes}}, \bibinfo {author} {\bibfnamefont {J.}~\bibnamefont {Liu}}, \bibinfo {author} {\bibfnamefont {C.-Y.}\ \bibnamefont {Lu}}, \bibinfo {author} {\bibfnamefont {N.}~\bibnamefont {Matsuda}}, \emph {et~al.},\ }\bibfield  {title} {\bibinfo {title} {The potential and global outlook of integrated photonics for quantum technologies},\ }\href@noop {} {\bibfield  {journal} {\bibinfo  {journal} {Nature Reviews Physics}\ }\textbf {\bibinfo {volume} {4}},\ \bibinfo {pages} {194} (\bibinfo {year} {2022})}\BibitemShut {NoStop}%
\bibitem [{\citenamefont {Ezratty}(2024)}]{Ezratty24}%
  \BibitemOpen
  \bibfield  {author} {\bibinfo {author} {\bibfnamefont {O.}~\bibnamefont {Ezratty}},\ }\href {https://arxiv.org/abs/2111.15352} {\bibinfo {title} {Understanding quantum technologies 2024}} (\bibinfo {year} {2024}),\ \Eprint {https://arxiv.org/abs/2111.15352} {arXiv:2111.15352 [quant-ph]} \BibitemShut {NoStop}%
\bibitem [{\citenamefont {Portmann}\ and\ \citenamefont {Renner}(2022)}]{Portmann22}%
  \BibitemOpen
  \bibfield  {author} {\bibinfo {author} {\bibfnamefont {C.}~\bibnamefont {Portmann}}\ and\ \bibinfo {author} {\bibfnamefont {R.}~\bibnamefont {Renner}},\ }\bibfield  {title} {\bibinfo {title} {Security in quantum cryptography},\ }\href {https://doi.org/10.1103/RevModPhys.94.025008} {\bibfield  {journal} {\bibinfo  {journal} {Rev. Mod. Phys.}\ }\textbf {\bibinfo {volume} {94}},\ \bibinfo {pages} {025008} (\bibinfo {year} {2022})}\BibitemShut {NoStop}%
\bibitem [{\citenamefont {Li}\ \emph {et~al.}(2024)\citenamefont {Li}, \citenamefont {Sun}, \citenamefont {Liu}, \citenamefont {Li},\ and\ \citenamefont {Wu}}]{li2024flying}%
  \BibitemOpen
  \bibfield  {author} {\bibinfo {author} {\bibfnamefont {W.}~\bibnamefont {Li}}, \bibinfo {author} {\bibfnamefont {H.}~\bibnamefont {Sun}}, \bibinfo {author} {\bibfnamefont {Y.}~\bibnamefont {Liu}}, \bibinfo {author} {\bibfnamefont {T.}~\bibnamefont {Li}},\ and\ \bibinfo {author} {\bibfnamefont {R.-B.}\ \bibnamefont {Wu}},\ }\bibfield  {title} {\bibinfo {title} {Flying-qubit control: Basics, progress, and outlook},\ }\href@noop {} {\bibfield  {journal} {\bibinfo  {journal} {Advanced Devices \& Instrumentation}\ }\textbf {\bibinfo {volume} {5}},\ \bibinfo {pages} {0059} (\bibinfo {year} {2024})}\BibitemShut {NoStop}%
\bibitem [{\citenamefont {Bhattacharjee}\ and\ \citenamefont {Dutta}(2021)}]{Bhattacharjee21}%
  \BibitemOpen
  \bibfield  {author} {\bibinfo {author} {\bibfnamefont {S.}~\bibnamefont {Bhattacharjee}}\ and\ \bibinfo {author} {\bibfnamefont {A.}~\bibnamefont {Dutta}},\ }\bibfield  {title} {\bibinfo {title} {Quantum thermal machines and batteries},\ }\href {https://doi.org/10.1140/epjb/s10051-021-00235-3} {\bibfield  {journal} {\bibinfo  {journal} {The European Physical Journal B}\ }\textbf {\bibinfo {volume} {94}},\ \bibinfo {pages} {239} (\bibinfo {year} {2021})}\BibitemShut {NoStop}%
\bibitem [{\citenamefont {Quach}\ \emph {et~al.}(2023)\citenamefont {Quach}, \citenamefont {Cerullo},\ and\ \citenamefont {Virgili}}]{Quach23}%
  \BibitemOpen
  \bibfield  {author} {\bibinfo {author} {\bibfnamefont {J.}~\bibnamefont {Quach}}, \bibinfo {author} {\bibfnamefont {G.}~\bibnamefont {Cerullo}},\ and\ \bibinfo {author} {\bibfnamefont {T.}~\bibnamefont {Virgili}},\ }\bibfield  {title} {\bibinfo {title} {Quantum batteries: The future of energy storage?},\ }\href {https://doi.org/https://doi.org/10.1016/j.joule.2023.09.003} {\bibfield  {journal} {\bibinfo  {journal} {Joule}\ }\textbf {\bibinfo {volume} {7}},\ \bibinfo {pages} {2195} (\bibinfo {year} {2023})}\BibitemShut {NoStop}%
\bibitem [{\citenamefont {Campaioli}\ \emph {et~al.}(2024)\citenamefont {Campaioli}, \citenamefont {Gherardini}, \citenamefont {Quach}, \citenamefont {Polini},\ and\ \citenamefont {Andolina}}]{campaioli2024colloquium}%
  \BibitemOpen
  \bibfield  {author} {\bibinfo {author} {\bibfnamefont {F.}~\bibnamefont {Campaioli}}, \bibinfo {author} {\bibfnamefont {S.}~\bibnamefont {Gherardini}}, \bibinfo {author} {\bibfnamefont {J.~Q.}\ \bibnamefont {Quach}}, \bibinfo {author} {\bibfnamefont {M.}~\bibnamefont {Polini}},\ and\ \bibinfo {author} {\bibfnamefont {G.~M.}\ \bibnamefont {Andolina}},\ }\bibfield  {title} {\bibinfo {title} {Colloquium: quantum batteries},\ }\href@noop {} {\bibfield  {journal} {\bibinfo  {journal} {Reviews of Modern Physics}\ }\textbf {\bibinfo {volume} {96}},\ \bibinfo {pages} {031001} (\bibinfo {year} {2024})}\BibitemShut {NoStop}%
\bibitem [{\citenamefont {Chiribella}\ \emph {et~al.}(2021)\citenamefont {Chiribella}, \citenamefont {Yang},\ and\ \citenamefont {Renner}}]{Chiribella21}%
  \BibitemOpen
  \bibfield  {author} {\bibinfo {author} {\bibfnamefont {G.}~\bibnamefont {Chiribella}}, \bibinfo {author} {\bibfnamefont {Y.}~\bibnamefont {Yang}},\ and\ \bibinfo {author} {\bibfnamefont {R.}~\bibnamefont {Renner}},\ }\bibfield  {title} {\bibinfo {title} {Fundamental energy requirement of reversible quantum operations},\ }\href {https://doi.org/10.1103/PhysRevX.11.021014} {\bibfield  {journal} {\bibinfo  {journal} {Phys. Rev. X}\ }\textbf {\bibinfo {volume} {11}},\ \bibinfo {pages} {021014} (\bibinfo {year} {2021})}\BibitemShut {NoStop}%
\bibitem [{\citenamefont {Cioni}\ \emph {et~al.}(2024)\citenamefont {Cioni}, \citenamefont {Menta}, \citenamefont {Aiudi}, \citenamefont {Polini},\ and\ \citenamefont {Giovannetti}}]{Cioni24}%
  \BibitemOpen
  \bibfield  {author} {\bibinfo {author} {\bibfnamefont {F.}~\bibnamefont {Cioni}}, \bibinfo {author} {\bibfnamefont {R.}~\bibnamefont {Menta}}, \bibinfo {author} {\bibfnamefont {R.}~\bibnamefont {Aiudi}}, \bibinfo {author} {\bibfnamefont {M.}~\bibnamefont {Polini}},\ and\ \bibinfo {author} {\bibfnamefont {V.}~\bibnamefont {Giovannetti}},\ }\href {https://arxiv.org/abs/2412.11782} {\bibinfo {title} {Conveyor-belt superconducting quantum computer}} (\bibinfo {year} {2024}),\ \Eprint {https://arxiv.org/abs/2412.11782} {arXiv:2412.11782 [quant-ph]} \BibitemShut {NoStop}%
\bibitem [{\citenamefont {Kurman}\ \emph {et~al.}(2025)\citenamefont {Kurman}, \citenamefont {Hymas}, \citenamefont {Fedorov}, \citenamefont {Munro},\ and\ \citenamefont {Quach}}]{Kurman25}%
  \BibitemOpen
  \bibfield  {author} {\bibinfo {author} {\bibfnamefont {Y.}~\bibnamefont {Kurman}}, \bibinfo {author} {\bibfnamefont {K.}~\bibnamefont {Hymas}}, \bibinfo {author} {\bibfnamefont {A.}~\bibnamefont {Fedorov}}, \bibinfo {author} {\bibfnamefont {W.~J.}\ \bibnamefont {Munro}},\ and\ \bibinfo {author} {\bibfnamefont {J.}~\bibnamefont {Quach}},\ }\href {https://arxiv.org/abs/2503.23610} {\bibinfo {title} {Quantum computation with quantum batteries}} (\bibinfo {year} {2025}),\ \Eprint {https://arxiv.org/abs/2503.23610} {arXiv:2503.23610 [quant-ph]} \BibitemShut {NoStop}%
\bibitem [{\citenamefont {Ferraro}\ \emph {et~al.}(2018)\citenamefont {Ferraro}, \citenamefont {Campisi}, \citenamefont {Andolina}, \citenamefont {Pellegrini},\ and\ \citenamefont {Polini}}]{Ferraro18}%
  \BibitemOpen
  \bibfield  {author} {\bibinfo {author} {\bibfnamefont {D.}~\bibnamefont {Ferraro}}, \bibinfo {author} {\bibfnamefont {M.}~\bibnamefont {Campisi}}, \bibinfo {author} {\bibfnamefont {G.~M.}\ \bibnamefont {Andolina}}, \bibinfo {author} {\bibfnamefont {V.}~\bibnamefont {Pellegrini}},\ and\ \bibinfo {author} {\bibfnamefont {M.}~\bibnamefont {Polini}},\ }\bibfield  {title} {\bibinfo {title} {High-power collective charging of a solid-state quantum battery},\ }\href {https://doi.org/10.1103/PhysRevLett.120.117702} {\bibfield  {journal} {\bibinfo  {journal} {Phys. Rev. Lett.}\ }\textbf {\bibinfo {volume} {120}},\ \bibinfo {pages} {117702} (\bibinfo {year} {2018})}\BibitemShut {NoStop}%
\bibitem [{\citenamefont {Ferraro}\ \emph {et~al.}(2019)\citenamefont {Ferraro}, \citenamefont {Andolina}, \citenamefont {Campisi}, \citenamefont {Pellegrini},\ and\ \citenamefont {Polini}}]{Ferraro19}%
  \BibitemOpen
  \bibfield  {author} {\bibinfo {author} {\bibfnamefont {D.}~\bibnamefont {Ferraro}}, \bibinfo {author} {\bibfnamefont {G.~M.}\ \bibnamefont {Andolina}}, \bibinfo {author} {\bibfnamefont {M.}~\bibnamefont {Campisi}}, \bibinfo {author} {\bibfnamefont {V.}~\bibnamefont {Pellegrini}},\ and\ \bibinfo {author} {\bibfnamefont {M.}~\bibnamefont {Polini}},\ }\bibfield  {title} {\bibinfo {title} {Quantum supercapacitors},\ }\href {https://doi.org/10.1103/PhysRevB.100.075433} {\bibfield  {journal} {\bibinfo  {journal} {Phys. Rev. B}\ }\textbf {\bibinfo {volume} {100}},\ \bibinfo {pages} {075433} (\bibinfo {year} {2019})}\BibitemShut {NoStop}%
\bibitem [{\citenamefont {Quach}\ \emph {et~al.}(2022)\citenamefont {Quach}, \citenamefont {McGhee}, \citenamefont {Ganzer}, \citenamefont {Rouse}, \citenamefont {Lovett}, \citenamefont {Gauger}, \citenamefont {Keeling}, \citenamefont {Cerullo}, \citenamefont {Lidzey},\ and\ \citenamefont {Virgili}}]{Quach22}%
  \BibitemOpen
  \bibfield  {author} {\bibinfo {author} {\bibfnamefont {J.~Q.}\ \bibnamefont {Quach}}, \bibinfo {author} {\bibfnamefont {K.~E.}\ \bibnamefont {McGhee}}, \bibinfo {author} {\bibfnamefont {L.}~\bibnamefont {Ganzer}}, \bibinfo {author} {\bibfnamefont {D.~M.}\ \bibnamefont {Rouse}}, \bibinfo {author} {\bibfnamefont {B.~W.}\ \bibnamefont {Lovett}}, \bibinfo {author} {\bibfnamefont {E.~M.}\ \bibnamefont {Gauger}}, \bibinfo {author} {\bibfnamefont {J.}~\bibnamefont {Keeling}}, \bibinfo {author} {\bibfnamefont {G.}~\bibnamefont {Cerullo}}, \bibinfo {author} {\bibfnamefont {D.~G.}\ \bibnamefont {Lidzey}},\ and\ \bibinfo {author} {\bibfnamefont {T.}~\bibnamefont {Virgili}},\ }\bibfield  {title} {\bibinfo {title} {Superabsorption in an organic microcavity: Toward a quantum battery},\ }\href {https://doi.org/10.1126/sciadv.abk3160} {\bibfield  {journal} {\bibinfo  {journal} {Science Advances}\ }\textbf {\bibinfo {volume} {8}},\ \bibinfo {pages} {eabk3160} (\bibinfo {year} {2022})},\ \Eprint
  {https://arxiv.org/abs/https://www.science.org/doi/pdf/10.1126/sciadv.abk3160} {https://www.science.org/doi/pdf/10.1126/sciadv.abk3160} \BibitemShut {NoStop}%
\bibitem [{\citenamefont {Gemme}\ \emph {et~al.}(2023)\citenamefont {Gemme}, \citenamefont {Andolina}, \citenamefont {Pellegrino}, \citenamefont {Sassetti},\ and\ \citenamefont {Ferraro}}]{Gemme23}%
  \BibitemOpen
  \bibfield  {author} {\bibinfo {author} {\bibfnamefont {G.}~\bibnamefont {Gemme}}, \bibinfo {author} {\bibfnamefont {G.~M.}\ \bibnamefont {Andolina}}, \bibinfo {author} {\bibfnamefont {F.~M.~D.}\ \bibnamefont {Pellegrino}}, \bibinfo {author} {\bibfnamefont {M.}~\bibnamefont {Sassetti}},\ and\ \bibinfo {author} {\bibfnamefont {D.}~\bibnamefont {Ferraro}},\ }\bibfield  {title} {\bibinfo {title} {Off-resonant dicke quantum battery: Charging by virtual photons},\ }\bibfield  {journal} {\bibinfo  {journal} {Batteries}\ }\textbf {\bibinfo {volume} {9}},\ \href {https://doi.org/10.3390/batteries9040197} {10.3390/batteries9040197} (\bibinfo {year} {2023})\BibitemShut {NoStop}%
\bibitem [{\citenamefont {Canzio}\ \emph {et~al.}(2025)\citenamefont {Canzio}, \citenamefont {Cavina}, \citenamefont {Polini},\ and\ \citenamefont {Giovannetti}}]{Canzio25}%
  \BibitemOpen
  \bibfield  {author} {\bibinfo {author} {\bibfnamefont {A.}~\bibnamefont {Canzio}}, \bibinfo {author} {\bibfnamefont {V.}~\bibnamefont {Cavina}}, \bibinfo {author} {\bibfnamefont {M.}~\bibnamefont {Polini}},\ and\ \bibinfo {author} {\bibfnamefont {V.}~\bibnamefont {Giovannetti}},\ }\bibfield  {title} {\bibinfo {title} {Single-atom dissipation and dephasing in dicke and tavis-cummings quantum batteries},\ }\href {https://doi.org/10.1103/PhysRevA.111.022222} {\bibfield  {journal} {\bibinfo  {journal} {Phys. Rev. A}\ }\textbf {\bibinfo {volume} {111}},\ \bibinfo {pages} {022222} (\bibinfo {year} {2025})}\BibitemShut {NoStop}%
\bibitem [{\citenamefont {Tibben}\ \emph {et~al.}(2024)\citenamefont {Tibben}, \citenamefont {Gaspera}, \citenamefont {van Embden}, \citenamefont {Reineck}, \citenamefont {Quach}, \citenamefont {Campaioli},\ and\ \citenamefont {Gómez}}]{Tibben24}%
  \BibitemOpen
  \bibfield  {author} {\bibinfo {author} {\bibfnamefont {D.~J.}\ \bibnamefont {Tibben}}, \bibinfo {author} {\bibfnamefont {E.~D.}\ \bibnamefont {Gaspera}}, \bibinfo {author} {\bibfnamefont {J.}~\bibnamefont {van Embden}}, \bibinfo {author} {\bibfnamefont {P.}~\bibnamefont {Reineck}}, \bibinfo {author} {\bibfnamefont {J.~Q.}\ \bibnamefont {Quach}}, \bibinfo {author} {\bibfnamefont {F.}~\bibnamefont {Campaioli}},\ and\ \bibinfo {author} {\bibfnamefont {D.~E.}\ \bibnamefont {Gómez}},\ }\href {https://arxiv.org/abs/2404.00198} {\bibinfo {title} {Extending the self-discharge time of dicke quantum batteries using molecular triplets}} (\bibinfo {year} {2024}),\ \Eprint {https://arxiv.org/abs/2404.00198} {arXiv:2404.00198 [quant-ph]} \BibitemShut {NoStop}%
\bibitem [{\citenamefont {Hymas}\ \emph {et~al.}(2025)\citenamefont {Hymas}, \citenamefont {Muir}, \citenamefont {Tibben}, \citenamefont {van Embden}, \citenamefont {Hirai}, \citenamefont {Dunn}, \citenamefont {Gómez}, \citenamefont {Hutchison}, \citenamefont {Smith},\ and\ \citenamefont {Quach}}]{Hymas25}%
  \BibitemOpen
  \bibfield  {author} {\bibinfo {author} {\bibfnamefont {K.}~\bibnamefont {Hymas}}, \bibinfo {author} {\bibfnamefont {J.~B.}\ \bibnamefont {Muir}}, \bibinfo {author} {\bibfnamefont {D.}~\bibnamefont {Tibben}}, \bibinfo {author} {\bibfnamefont {J.}~\bibnamefont {van Embden}}, \bibinfo {author} {\bibfnamefont {T.}~\bibnamefont {Hirai}}, \bibinfo {author} {\bibfnamefont {C.~J.}\ \bibnamefont {Dunn}}, \bibinfo {author} {\bibfnamefont {D.~E.}\ \bibnamefont {Gómez}}, \bibinfo {author} {\bibfnamefont {J.~A.}\ \bibnamefont {Hutchison}}, \bibinfo {author} {\bibfnamefont {T.~A.}\ \bibnamefont {Smith}},\ and\ \bibinfo {author} {\bibfnamefont {J.~Q.}\ \bibnamefont {Quach}},\ }\href {https://arxiv.org/abs/2501.16541} {\bibinfo {title} {Experimental demonstration of a scalable room-temperature quantum battery}} (\bibinfo {year} {2025}),\ \Eprint {https://arxiv.org/abs/2501.16541} {arXiv:2501.16541 [quant-ph]} \BibitemShut {NoStop}%
\bibitem [{\citenamefont {Seah}\ \emph {et~al.}(2021)\citenamefont {Seah}, \citenamefont {Perarnau-Llobet}, \citenamefont {Haack}, \citenamefont {Brunner},\ and\ \citenamefont {Nimmrichter}}]{Seah21}%
  \BibitemOpen
  \bibfield  {author} {\bibinfo {author} {\bibfnamefont {S.}~\bibnamefont {Seah}}, \bibinfo {author} {\bibfnamefont {M.}~\bibnamefont {Perarnau-Llobet}}, \bibinfo {author} {\bibfnamefont {G.}~\bibnamefont {Haack}}, \bibinfo {author} {\bibfnamefont {N.}~\bibnamefont {Brunner}},\ and\ \bibinfo {author} {\bibfnamefont {S.}~\bibnamefont {Nimmrichter}},\ }\bibfield  {title} {\bibinfo {title} {Quantum speed-up in collisional battery charging},\ }\href {https://doi.org/10.1103/PhysRevLett.127.100601} {\bibfield  {journal} {\bibinfo  {journal} {Phys. Rev. Lett.}\ }\textbf {\bibinfo {volume} {127}},\ \bibinfo {pages} {100601} (\bibinfo {year} {2021})}\BibitemShut {NoStop}%
\bibitem [{\citenamefont {Shaghaghi}\ \emph {et~al.}(2022)\citenamefont {Shaghaghi}, \citenamefont {Singh}, \citenamefont {Benenti},\ and\ \citenamefont {Rosa}}]{Shaghaghi22}%
  \BibitemOpen
  \bibfield  {author} {\bibinfo {author} {\bibfnamefont {V.}~\bibnamefont {Shaghaghi}}, \bibinfo {author} {\bibfnamefont {V.}~\bibnamefont {Singh}}, \bibinfo {author} {\bibfnamefont {G.}~\bibnamefont {Benenti}},\ and\ \bibinfo {author} {\bibfnamefont {D.}~\bibnamefont {Rosa}},\ }\bibfield  {title} {\bibinfo {title} {Micromasers as quantum batteries},\ }\href {https://doi.org/10.1088/2058-9565/ac8829} {\bibfield  {journal} {\bibinfo  {journal} {Quantum Science and Technology}\ }\textbf {\bibinfo {volume} {7}},\ \bibinfo {pages} {04LT01} (\bibinfo {year} {2022})}\BibitemShut {NoStop}%
\bibitem [{\citenamefont {Shaghaghi}\ \emph {et~al.}(2023)\citenamefont {Shaghaghi}, \citenamefont {Singh}, \citenamefont {Carrega}, \citenamefont {Rosa},\ and\ \citenamefont {Benenti}}]{Shaghaghi23}%
  \BibitemOpen
  \bibfield  {author} {\bibinfo {author} {\bibfnamefont {V.}~\bibnamefont {Shaghaghi}}, \bibinfo {author} {\bibfnamefont {V.}~\bibnamefont {Singh}}, \bibinfo {author} {\bibfnamefont {M.}~\bibnamefont {Carrega}}, \bibinfo {author} {\bibfnamefont {D.}~\bibnamefont {Rosa}},\ and\ \bibinfo {author} {\bibfnamefont {G.}~\bibnamefont {Benenti}},\ }\bibfield  {title} {\bibinfo {title} {Lossy micromaser battery: Almost pure states in the jaynes–cummings regime},\ }\bibfield  {journal} {\bibinfo  {journal} {Entropy}\ }\textbf {\bibinfo {volume} {25}},\ \href {https://doi.org/10.3390/e25030430} {10.3390/e25030430} (\bibinfo {year} {2023})\BibitemShut {NoStop}%
\bibitem [{\citenamefont {Rodr\'{\i}guez}\ \emph {et~al.}(2023)\citenamefont {Rodr\'{\i}guez}, \citenamefont {Rosa},\ and\ \citenamefont {Olle}}]{Rodriguez23}%
  \BibitemOpen
  \bibfield  {author} {\bibinfo {author} {\bibfnamefont {C.}~\bibnamefont {Rodr\'{\i}guez}}, \bibinfo {author} {\bibfnamefont {D.}~\bibnamefont {Rosa}},\ and\ \bibinfo {author} {\bibfnamefont {J.}~\bibnamefont {Olle}},\ }\bibfield  {title} {\bibinfo {title} {Artificial intelligence discovery of a charging protocol in a micromaser quantum battery},\ }\href {https://doi.org/10.1103/PhysRevA.108.042618} {\bibfield  {journal} {\bibinfo  {journal} {Phys. Rev. A}\ }\textbf {\bibinfo {volume} {108}},\ \bibinfo {pages} {042618} (\bibinfo {year} {2023})}\BibitemShut {NoStop}%
\bibitem [{\citenamefont {Downing}\ and\ \citenamefont {Ukhtary}(2024)}]{Downing24}%
  \BibitemOpen
  \bibfield  {author} {\bibinfo {author} {\bibfnamefont {C.}~\bibnamefont {Downing}}\ and\ \bibinfo {author} {\bibfnamefont {M.}~\bibnamefont {Ukhtary}},\ }\bibfield  {title} {\bibinfo {title} {Two-photon charging of a quantum battery with a gaussian pulse envelope},\ }\href {https://doi.org/https://doi.org/10.1016/j.physleta.2024.129693} {\bibfield  {journal} {\bibinfo  {journal} {Physics Letters A}\ }\textbf {\bibinfo {volume} {518}},\ \bibinfo {pages} {129693} (\bibinfo {year} {2024})}\BibitemShut {NoStop}%
\bibitem [{\citenamefont {Andolina}\ \emph {et~al.}(2025)\citenamefont {Andolina}, \citenamefont {Stanzione}, \citenamefont {Giovannetti},\ and\ \citenamefont {Polini}}]{Andolina25}%
  \BibitemOpen
  \bibfield  {author} {\bibinfo {author} {\bibfnamefont {G.~M.}\ \bibnamefont {Andolina}}, \bibinfo {author} {\bibfnamefont {V.}~\bibnamefont {Stanzione}}, \bibinfo {author} {\bibfnamefont {V.}~\bibnamefont {Giovannetti}},\ and\ \bibinfo {author} {\bibfnamefont {M.}~\bibnamefont {Polini}},\ }\bibfield  {title} {\bibinfo {title} {Genuine quantum advantage in anharmonic bosonic quantum batteries},\ }\href {https://doi.org/10.1103/kzvn-dj7v} {\bibfield  {journal} {\bibinfo  {journal} {Phys. Rev. Lett.}\ }\textbf {\bibinfo {volume} {134}},\ \bibinfo {pages} {240403} (\bibinfo {year} {2025})}\BibitemShut {NoStop}%
\bibitem [{\citenamefont {Kirton}\ \emph {et~al.}(2019)\citenamefont {Kirton}, \citenamefont {Roses}, \citenamefont {Keeling},\ and\ \citenamefont {Dalla~Torre}}]{Kirton19}%
  \BibitemOpen
  \bibfield  {author} {\bibinfo {author} {\bibfnamefont {P.}~\bibnamefont {Kirton}}, \bibinfo {author} {\bibfnamefont {M.~M.}\ \bibnamefont {Roses}}, \bibinfo {author} {\bibfnamefont {J.}~\bibnamefont {Keeling}},\ and\ \bibinfo {author} {\bibfnamefont {E.~G.}\ \bibnamefont {Dalla~Torre}},\ }\bibfield  {title} {\bibinfo {title} {Introduction to the dicke model: From equilibrium to nonequilibrium, and vice versa},\ }\href {https://doi.org/https://doi.org/10.1002/qute.201800043} {\bibfield  {journal} {\bibinfo  {journal} {Advanced Quantum Technologies}\ }\textbf {\bibinfo {volume} {2}},\ \bibinfo {pages} {1800043} (\bibinfo {year} {2019})}\BibitemShut {NoStop}%
\bibitem [{\citenamefont {Haroche}(2013)}]{Haroche13}%
  \BibitemOpen
  \bibfield  {author} {\bibinfo {author} {\bibfnamefont {S.}~\bibnamefont {Haroche}},\ }\bibfield  {title} {\bibinfo {title} {Nobel lecture: Controlling photons in a box and exploring the quantum to classical boundary},\ }\href {https://doi.org/10.1103/RevModPhys.85.1083} {\bibfield  {journal} {\bibinfo  {journal} {Rev. Mod. Phys.}\ }\textbf {\bibinfo {volume} {85}},\ \bibinfo {pages} {1083} (\bibinfo {year} {2013})}\BibitemShut {NoStop}%
\bibitem [{\citenamefont {Olaya-Castro}\ \emph {et~al.}(2008)\citenamefont {Olaya-Castro}, \citenamefont {Lee}, \citenamefont {Olsen},\ and\ \citenamefont {Johnson}}]{Olaya-Castro08}%
  \BibitemOpen
  \bibfield  {author} {\bibinfo {author} {\bibfnamefont {A.}~\bibnamefont {Olaya-Castro}}, \bibinfo {author} {\bibfnamefont {C.~F.}\ \bibnamefont {Lee}}, \bibinfo {author} {\bibfnamefont {F.~F.}\ \bibnamefont {Olsen}},\ and\ \bibinfo {author} {\bibfnamefont {N.~F.}\ \bibnamefont {Johnson}},\ }\bibfield  {title} {\bibinfo {title} {Efficiency of energy transfer in a light-harvesting system under quantum coherence},\ }\href {https://doi.org/10.1103/PhysRevB.78.085115} {\bibfield  {journal} {\bibinfo  {journal} {Phys. Rev. B}\ }\textbf {\bibinfo {volume} {78}},\ \bibinfo {pages} {085115} (\bibinfo {year} {2008})}\BibitemShut {NoStop}%
\bibitem [{\citenamefont {Sahoo}(2011)}]{Sahoo11}%
  \BibitemOpen
  \bibfield  {author} {\bibinfo {author} {\bibfnamefont {H.}~\bibnamefont {Sahoo}},\ }\bibfield  {title} {\bibinfo {title} {Förster resonance energy transfer – a spectroscopic nanoruler: Principle and applications},\ }\href {https://doi.org/https://doi.org/10.1016/j.jphotochemrev.2011.05.001} {\bibfield  {journal} {\bibinfo  {journal} {Journal of Photochemistry and Photobiology C: Photochemistry Reviews}\ }\textbf {\bibinfo {volume} {12}},\ \bibinfo {pages} {20} (\bibinfo {year} {2011})}\BibitemShut {NoStop}%
\bibitem [{\citenamefont {Sillanp{\"a}{\"a}}\ \emph {et~al.}(2007)\citenamefont {Sillanp{\"a}{\"a}}, \citenamefont {Park},\ and\ \citenamefont {Simmonds}}]{Sillanpää07}%
  \BibitemOpen
  \bibfield  {author} {\bibinfo {author} {\bibfnamefont {M.~A.}\ \bibnamefont {Sillanp{\"a}{\"a}}}, \bibinfo {author} {\bibfnamefont {J.~I.}\ \bibnamefont {Park}},\ and\ \bibinfo {author} {\bibfnamefont {R.~W.}\ \bibnamefont {Simmonds}},\ }\bibfield  {title} {\bibinfo {title} {Coherent quantum state storage and transfer between two phase qubits via a resonant cavity},\ }\href {https://doi.org/10.1038/nature06124} {\bibfield  {journal} {\bibinfo  {journal} {Nature}\ }\textbf {\bibinfo {volume} {449}},\ \bibinfo {pages} {438} (\bibinfo {year} {2007})}\BibitemShut {NoStop}%
\bibitem [{\citenamefont {Crescente}\ \emph {et~al.}(2022)\citenamefont {Crescente}, \citenamefont {Ferraro}, \citenamefont {Carrega},\ and\ \citenamefont {Sassetti}}]{Crescente22}%
  \BibitemOpen
  \bibfield  {author} {\bibinfo {author} {\bibfnamefont {A.}~\bibnamefont {Crescente}}, \bibinfo {author} {\bibfnamefont {D.}~\bibnamefont {Ferraro}}, \bibinfo {author} {\bibfnamefont {M.}~\bibnamefont {Carrega}},\ and\ \bibinfo {author} {\bibfnamefont {M.}~\bibnamefont {Sassetti}},\ }\bibfield  {title} {\bibinfo {title} {Enhancing coherent energy transfer between quantum devices via a mediator},\ }\href {https://doi.org/10.1103/PhysRevResearch.4.033216} {\bibfield  {journal} {\bibinfo  {journal} {Phys. Rev. Res.}\ }\textbf {\bibinfo {volume} {4}},\ \bibinfo {pages} {033216} (\bibinfo {year} {2022})}\BibitemShut {NoStop}%
\bibitem [{\citenamefont {Crescente}\ \emph {et~al.}(2024)\citenamefont {Crescente}, \citenamefont {Ferraro},\ and\ \citenamefont {Sassetti}}]{Crescente24}%
  \BibitemOpen
  \bibfield  {author} {\bibinfo {author} {\bibfnamefont {A.}~\bibnamefont {Crescente}}, \bibinfo {author} {\bibfnamefont {D.}~\bibnamefont {Ferraro}},\ and\ \bibinfo {author} {\bibfnamefont {M.}~\bibnamefont {Sassetti}},\ }\bibfield  {title} {\bibinfo {title} {Boosting energy transfer between quantum devices through spectrum engineering in the dissipative ultrastrong coupling regime},\ }\href {https://doi.org/10.1103/PhysRevResearch.6.023092} {\bibfield  {journal} {\bibinfo  {journal} {Phys. Rev. Res.}\ }\textbf {\bibinfo {volume} {6}},\ \bibinfo {pages} {023092} (\bibinfo {year} {2024})}\BibitemShut {NoStop}%
\bibitem [{\citenamefont {Walther}\ \emph {et~al.}(2006)\citenamefont {Walther}, \citenamefont {Varcoe}, \citenamefont {Englert},\ and\ \citenamefont {Becker}}]{walther2006cavity}%
  \BibitemOpen
  \bibfield  {author} {\bibinfo {author} {\bibfnamefont {H.}~\bibnamefont {Walther}}, \bibinfo {author} {\bibfnamefont {B.~T.}\ \bibnamefont {Varcoe}}, \bibinfo {author} {\bibfnamefont {B.-G.}\ \bibnamefont {Englert}},\ and\ \bibinfo {author} {\bibfnamefont {T.}~\bibnamefont {Becker}},\ }\bibfield  {title} {\bibinfo {title} {Cavity quantum electrodynamics},\ }\href@noop {} {\bibfield  {journal} {\bibinfo  {journal} {Reports on Progress in Physics}\ }\textbf {\bibinfo {volume} {69}},\ \bibinfo {pages} {1325} (\bibinfo {year} {2006})}\BibitemShut {NoStop}%
\bibitem [{\citenamefont {Christandl}\ \emph {et~al.}(2004)\citenamefont {Christandl}, \citenamefont {Datta}, \citenamefont {Ekert},\ and\ \citenamefont {Landahl}}]{christandl2004perfect}%
  \BibitemOpen
  \bibfield  {author} {\bibinfo {author} {\bibfnamefont {M.}~\bibnamefont {Christandl}}, \bibinfo {author} {\bibfnamefont {N.}~\bibnamefont {Datta}}, \bibinfo {author} {\bibfnamefont {A.}~\bibnamefont {Ekert}},\ and\ \bibinfo {author} {\bibfnamefont {A.~J.}\ \bibnamefont {Landahl}},\ }\bibfield  {title} {\bibinfo {title} {Perfect state transfer in quantum spin networks},\ }\href@noop {} {\bibfield  {journal} {\bibinfo  {journal} {Physical review letters}\ }\textbf {\bibinfo {volume} {92}},\ \bibinfo {pages} {187902} (\bibinfo {year} {2004})}\BibitemShut {NoStop}%
\bibitem [{\citenamefont {Allahverdyan}\ \emph {et~al.}(2004)\citenamefont {Allahverdyan}, \citenamefont {Balian},\ and\ \citenamefont {Nieuwenhuizen}}]{allahverdyan2004maximal}%
  \BibitemOpen
  \bibfield  {author} {\bibinfo {author} {\bibfnamefont {A.~E.}\ \bibnamefont {Allahverdyan}}, \bibinfo {author} {\bibfnamefont {R.}~\bibnamefont {Balian}},\ and\ \bibinfo {author} {\bibfnamefont {T.~M.}\ \bibnamefont {Nieuwenhuizen}},\ }\bibfield  {title} {\bibinfo {title} {Maximal work extraction from finite quantum systems},\ }\href@noop {} {\bibfield  {journal} {\bibinfo  {journal} {Europhysics Letters}\ }\textbf {\bibinfo {volume} {67}},\ \bibinfo {pages} {565} (\bibinfo {year} {2004})}\BibitemShut {NoStop}%
\bibitem [{\citenamefont {Alicki}\ and\ \citenamefont {Fannes}(2013)}]{alicki2013entanglement}%
  \BibitemOpen
  \bibfield  {author} {\bibinfo {author} {\bibfnamefont {R.}~\bibnamefont {Alicki}}\ and\ \bibinfo {author} {\bibfnamefont {M.}~\bibnamefont {Fannes}},\ }\bibfield  {title} {\bibinfo {title} {Entanglement boost for extractable work from ensembles of quantum batteries},\ }\href@noop {} {\bibfield  {journal} {\bibinfo  {journal} {Physical Review E—Statistical, Nonlinear, and Soft Matter Physics}\ }\textbf {\bibinfo {volume} {87}},\ \bibinfo {pages} {042123} (\bibinfo {year} {2013})}\BibitemShut {NoStop}%
\bibitem [{\citenamefont {Touil}\ \emph {et~al.}(2021)\citenamefont {Touil}, \citenamefont {{\c{C}}akmak},\ and\ \citenamefont {Deffner}}]{touil2021ergotropy}%
  \BibitemOpen
  \bibfield  {author} {\bibinfo {author} {\bibfnamefont {A.}~\bibnamefont {Touil}}, \bibinfo {author} {\bibfnamefont {B.}~\bibnamefont {{\c{C}}akmak}},\ and\ \bibinfo {author} {\bibfnamefont {S.}~\bibnamefont {Deffner}},\ }\bibfield  {title} {\bibinfo {title} {Ergotropy from quantum and classical correlations},\ }\href@noop {} {\bibfield  {journal} {\bibinfo  {journal} {Journal of Physics A: Mathematical and Theoretical}\ }\textbf {\bibinfo {volume} {55}},\ \bibinfo {pages} {025301} (\bibinfo {year} {2021})}\BibitemShut {NoStop}%
\bibitem [{\citenamefont {Andolina}\ \emph {et~al.}(2018)\citenamefont {Andolina}, \citenamefont {Farina}, \citenamefont {Mari}, \citenamefont {Pellegrini}, \citenamefont {Giovannetti},\ and\ \citenamefont {Polini}}]{andolina2018charger}%
  \BibitemOpen
  \bibfield  {author} {\bibinfo {author} {\bibfnamefont {G.~M.}\ \bibnamefont {Andolina}}, \bibinfo {author} {\bibfnamefont {D.}~\bibnamefont {Farina}}, \bibinfo {author} {\bibfnamefont {A.}~\bibnamefont {Mari}}, \bibinfo {author} {\bibfnamefont {V.}~\bibnamefont {Pellegrini}}, \bibinfo {author} {\bibfnamefont {V.}~\bibnamefont {Giovannetti}},\ and\ \bibinfo {author} {\bibfnamefont {M.}~\bibnamefont {Polini}},\ }\bibfield  {title} {\bibinfo {title} {Charger-mediated energy transfer in exactly solvable models for quantum batteries},\ }\href {https://doi.org/10.1103/PhysRevB.98.205423} {\bibfield  {journal} {\bibinfo  {journal} {Phys. Rev. B}\ }\textbf {\bibinfo {volume} {98}},\ \bibinfo {pages} {205423} (\bibinfo {year} {2018})}\BibitemShut {NoStop}%
\bibitem [{\citenamefont {Cavaliere}\ \emph {et~al.}(2025)\citenamefont {Cavaliere}, \citenamefont {Gemme}, \citenamefont {Benenti}, \citenamefont {Ferraro},\ and\ \citenamefont {Sassetti}}]{Cavaliere25}%
  \BibitemOpen
  \bibfield  {author} {\bibinfo {author} {\bibfnamefont {F.}~\bibnamefont {Cavaliere}}, \bibinfo {author} {\bibfnamefont {G.}~\bibnamefont {Gemme}}, \bibinfo {author} {\bibfnamefont {G.}~\bibnamefont {Benenti}}, \bibinfo {author} {\bibfnamefont {D.}~\bibnamefont {Ferraro}},\ and\ \bibinfo {author} {\bibfnamefont {M.}~\bibnamefont {Sassetti}},\ }\bibfield  {title} {\bibinfo {title} {Dynamical blockade of a reservoir for optimal performances of a quantum battery},\ }\href {https://doi.org/10.1038/s42005-025-01993-7} {\bibfield  {journal} {\bibinfo  {journal} {Communications Physics}\ }\textbf {\bibinfo {volume} {8}},\ \bibinfo {pages} {76} (\bibinfo {year} {2025})}\BibitemShut {NoStop}%
\bibitem [{\citenamefont {Morrone}\ \emph {et~al.}(2023)\citenamefont {Morrone}, \citenamefont {Rossi},\ and\ \citenamefont {Genoni}}]{Morone23}%
  \BibitemOpen
  \bibfield  {author} {\bibinfo {author} {\bibfnamefont {D.}~\bibnamefont {Morrone}}, \bibinfo {author} {\bibfnamefont {M.~A.}\ \bibnamefont {Rossi}},\ and\ \bibinfo {author} {\bibfnamefont {M.~G.}\ \bibnamefont {Genoni}},\ }\bibfield  {title} {\bibinfo {title} {Daemonic ergotropy in continuously monitored open quantum batteries},\ }\href {https://doi.org/10.1103/PhysRevApplied.20.044073} {\bibfield  {journal} {\bibinfo  {journal} {Phys. Rev. Appl.}\ }\textbf {\bibinfo {volume} {20}},\ \bibinfo {pages} {044073} (\bibinfo {year} {2023})}\BibitemShut {NoStop}%
\bibitem [{\citenamefont {Castellano}\ \emph {et~al.}(2024)\citenamefont {Castellano}, \citenamefont {Farina}, \citenamefont {Giovannetti},\ and\ \citenamefont {Acin}}]{Castellano24}%
  \BibitemOpen
  \bibfield  {author} {\bibinfo {author} {\bibfnamefont {R.}~\bibnamefont {Castellano}}, \bibinfo {author} {\bibfnamefont {D.}~\bibnamefont {Farina}}, \bibinfo {author} {\bibfnamefont {V.}~\bibnamefont {Giovannetti}},\ and\ \bibinfo {author} {\bibfnamefont {A.}~\bibnamefont {Acin}},\ }\bibfield  {title} {\bibinfo {title} {Extended local ergotropy},\ }\href {https://doi.org/10.1103/PhysRevLett.133.150402} {\bibfield  {journal} {\bibinfo  {journal} {Phys. Rev. Lett.}\ }\textbf {\bibinfo {volume} {133}},\ \bibinfo {pages} {150402} (\bibinfo {year} {2024})}\BibitemShut {NoStop}%
\bibitem [{\citenamefont {Navid~Elyasi}\ \emph {et~al.}(2025)\citenamefont {Navid~Elyasi}, \citenamefont {Rossi},\ and\ \citenamefont {Genoni}}]{Elyasi25}%
  \BibitemOpen
  \bibfield  {author} {\bibinfo {author} {\bibfnamefont {S.}~\bibnamefont {Navid~Elyasi}}, \bibinfo {author} {\bibfnamefont {M.~A.~C.}\ \bibnamefont {Rossi}},\ and\ \bibinfo {author} {\bibfnamefont {M.~G.}\ \bibnamefont {Genoni}},\ }\bibfield  {title} {\bibinfo {title} {Experimental simulation of daemonic work extraction in open quantum batteries on a digital quantum computer},\ }\href {https://doi.org/10.1088/2058-9565/adae2d} {\bibfield  {journal} {\bibinfo  {journal} {Quantum Science and Technology}\ }\textbf {\bibinfo {volume} {10}},\ \bibinfo {pages} {025017} (\bibinfo {year} {2025})}\BibitemShut {NoStop}%
\bibitem [{\citenamefont {Carrega}\ \emph {et~al.}(2016)\citenamefont {Carrega}, \citenamefont {Solinas}, \citenamefont {Sassetti},\ and\ \citenamefont {Weiss}}]{Carrega16}%
  \BibitemOpen
  \bibfield  {author} {\bibinfo {author} {\bibfnamefont {M.}~\bibnamefont {Carrega}}, \bibinfo {author} {\bibfnamefont {P.}~\bibnamefont {Solinas}}, \bibinfo {author} {\bibfnamefont {M.}~\bibnamefont {Sassetti}},\ and\ \bibinfo {author} {\bibfnamefont {U.}~\bibnamefont {Weiss}},\ }\bibfield  {title} {\bibinfo {title} {Energy exchange in driven open quantum systems at strong coupling},\ }\href {https://doi.org/10.1103/PhysRevLett.116.240403} {\bibfield  {journal} {\bibinfo  {journal} {Phys. Rev. Lett.}\ }\textbf {\bibinfo {volume} {116}},\ \bibinfo {pages} {240403} (\bibinfo {year} {2016})}\BibitemShut {NoStop}%
\bibitem [{\citenamefont {Andolina}\ \emph {et~al.}(2019)\citenamefont {Andolina}, \citenamefont {Keck}, \citenamefont {Mari}, \citenamefont {Campisi}, \citenamefont {Giovannetti},\ and\ \citenamefont {Polini}}]{Andolina19}%
  \BibitemOpen
  \bibfield  {author} {\bibinfo {author} {\bibfnamefont {G.~M.}\ \bibnamefont {Andolina}}, \bibinfo {author} {\bibfnamefont {M.}~\bibnamefont {Keck}}, \bibinfo {author} {\bibfnamefont {A.}~\bibnamefont {Mari}}, \bibinfo {author} {\bibfnamefont {M.}~\bibnamefont {Campisi}}, \bibinfo {author} {\bibfnamefont {V.}~\bibnamefont {Giovannetti}},\ and\ \bibinfo {author} {\bibfnamefont {M.}~\bibnamefont {Polini}},\ }\bibfield  {title} {\bibinfo {title} {Extractable work, the role of correlations, and asymptotic freedom in quantum batteries},\ }\href {https://doi.org/10.1103/PhysRevLett.122.047702} {\bibfield  {journal} {\bibinfo  {journal} {Phys. Rev. Lett.}\ }\textbf {\bibinfo {volume} {122}},\ \bibinfo {pages} {047702} (\bibinfo {year} {2019})}\BibitemShut {NoStop}%
\bibitem [{\citenamefont {Ma}\ \emph {et~al.}(2024)\citenamefont {Ma}, \citenamefont {Xu}, \citenamefont {Li}, \citenamefont {Li},\ and\ \citenamefont {Zhu}}]{ma2024enhancing}%
  \BibitemOpen
  \bibfield  {author} {\bibinfo {author} {\bibfnamefont {H.-B.}\ \bibnamefont {Ma}}, \bibinfo {author} {\bibfnamefont {K.}~\bibnamefont {Xu}}, \bibinfo {author} {\bibfnamefont {H.-G.}\ \bibnamefont {Li}}, \bibinfo {author} {\bibfnamefont {Z.-G.}\ \bibnamefont {Li}},\ and\ \bibinfo {author} {\bibfnamefont {H.-J.}\ \bibnamefont {Zhu}},\ }\bibfield  {title} {\bibinfo {title} {Enhancing the charging performance of quantum batteries with the work medium of an entangled coupled-cavity array},\ }\href@noop {} {\bibfield  {journal} {\bibinfo  {journal} {Physical Review A}\ }\textbf {\bibinfo {volume} {110}},\ \bibinfo {pages} {022433} (\bibinfo {year} {2024})}\BibitemShut {NoStop}%
\bibitem [{\citenamefont {D{\"u}r}\ \emph {et~al.}(2000)\citenamefont {D{\"u}r}, \citenamefont {Vidal},\ and\ \citenamefont {Cirac}}]{dur2000three}%
  \BibitemOpen
  \bibfield  {author} {\bibinfo {author} {\bibfnamefont {W.}~\bibnamefont {D{\"u}r}}, \bibinfo {author} {\bibfnamefont {G.}~\bibnamefont {Vidal}},\ and\ \bibinfo {author} {\bibfnamefont {J.~I.}\ \bibnamefont {Cirac}},\ }\bibfield  {title} {\bibinfo {title} {Three qubits can be entangled in two inequivalent ways},\ }\href@noop {} {\bibfield  {journal} {\bibinfo  {journal} {Physical Review A}\ }\textbf {\bibinfo {volume} {62}},\ \bibinfo {pages} {062314} (\bibinfo {year} {2000})}\BibitemShut {NoStop}%
\bibitem [{\citenamefont {Li}\ \emph {et~al.}(2018)\citenamefont {Li}, \citenamefont {Ma}, \citenamefont {Han}, \citenamefont {Chen}, \citenamefont {Xu}, \citenamefont {Cai}, \citenamefont {Wang}, \citenamefont {Song}, \citenamefont {Xue}, \citenamefont {Yin},\ and\ \citenamefont {Sun}}]{PhysRevApplied.10.054009}%
  \BibitemOpen
  \bibfield  {author} {\bibinfo {author} {\bibfnamefont {X.}~\bibnamefont {Li}}, \bibinfo {author} {\bibfnamefont {Y.}~\bibnamefont {Ma}}, \bibinfo {author} {\bibfnamefont {J.}~\bibnamefont {Han}}, \bibinfo {author} {\bibfnamefont {T.}~\bibnamefont {Chen}}, \bibinfo {author} {\bibfnamefont {Y.}~\bibnamefont {Xu}}, \bibinfo {author} {\bibfnamefont {W.}~\bibnamefont {Cai}}, \bibinfo {author} {\bibfnamefont {H.}~\bibnamefont {Wang}}, \bibinfo {author} {\bibfnamefont {Y.}~\bibnamefont {Song}}, \bibinfo {author} {\bibfnamefont {Z.-Y.}\ \bibnamefont {Xue}}, \bibinfo {author} {\bibfnamefont {Z.-q.}\ \bibnamefont {Yin}},\ and\ \bibinfo {author} {\bibfnamefont {L.}~\bibnamefont {Sun}},\ }\bibfield  {title} {\bibinfo {title} {Perfect quantum state transfer in a superconducting qubit chain with parametrically tunable couplings},\ }\href {https://doi.org/10.1103/PhysRevApplied.10.054009} {\bibfield  {journal} {\bibinfo  {journal} {Phys. Rev. Appl.}\ }\textbf {\bibinfo {volume} {10}},\ \bibinfo {pages} {054009} (\bibinfo
  {year} {2018})}\BibitemShut {NoStop}%
\bibitem [{\citenamefont {Chapman}\ \emph {et~al.}(2016)\citenamefont {Chapman}, \citenamefont {Santandrea}, \citenamefont {Huang}, \citenamefont {Corrielli}, \citenamefont {Crespi}, \citenamefont {Yung}, \citenamefont {Osellame},\ and\ \citenamefont {Peruzzo}}]{chapman2016experimental}%
  \BibitemOpen
  \bibfield  {author} {\bibinfo {author} {\bibfnamefont {R.~J.}\ \bibnamefont {Chapman}}, \bibinfo {author} {\bibfnamefont {M.}~\bibnamefont {Santandrea}}, \bibinfo {author} {\bibfnamefont {Z.}~\bibnamefont {Huang}}, \bibinfo {author} {\bibfnamefont {G.}~\bibnamefont {Corrielli}}, \bibinfo {author} {\bibfnamefont {A.}~\bibnamefont {Crespi}}, \bibinfo {author} {\bibfnamefont {M.-H.}\ \bibnamefont {Yung}}, \bibinfo {author} {\bibfnamefont {R.}~\bibnamefont {Osellame}},\ and\ \bibinfo {author} {\bibfnamefont {A.}~\bibnamefont {Peruzzo}},\ }\bibfield  {title} {\bibinfo {title} {Experimental perfect state transfer of an entangled photonic qubit},\ }\href@noop {} {\bibfield  {journal} {\bibinfo  {journal} {Nature communications}\ }\textbf {\bibinfo {volume} {7}},\ \bibinfo {pages} {11339} (\bibinfo {year} {2016})}\BibitemShut {NoStop}%
\bibitem [{\citenamefont {Le}\ \emph {et~al.}(2018)\citenamefont {Le}, \citenamefont {Levinsen}, \citenamefont {Modi}, \citenamefont {Parish},\ and\ \citenamefont {Pollock}}]{Le18}%
  \BibitemOpen
  \bibfield  {author} {\bibinfo {author} {\bibfnamefont {T.~P.}\ \bibnamefont {Le}}, \bibinfo {author} {\bibfnamefont {J.}~\bibnamefont {Levinsen}}, \bibinfo {author} {\bibfnamefont {K.}~\bibnamefont {Modi}}, \bibinfo {author} {\bibfnamefont {M.~M.}\ \bibnamefont {Parish}},\ and\ \bibinfo {author} {\bibfnamefont {F.~A.}\ \bibnamefont {Pollock}},\ }\bibfield  {title} {\bibinfo {title} {Spin-chain model of a many-body quantum battery},\ }\href {https://doi.org/10.1103/PhysRevA.97.022106} {\bibfield  {journal} {\bibinfo  {journal} {Phys. Rev. A}\ }\textbf {\bibinfo {volume} {97}},\ \bibinfo {pages} {022106} (\bibinfo {year} {2018})}\BibitemShut {NoStop}%
\bibitem [{\citenamefont {Zhao}\ \emph {et~al.}(2021)\citenamefont {Zhao}, \citenamefont {Dou},\ and\ \citenamefont {Zhao}}]{Zhao21}%
  \BibitemOpen
  \bibfield  {author} {\bibinfo {author} {\bibfnamefont {F.}~\bibnamefont {Zhao}}, \bibinfo {author} {\bibfnamefont {F.-Q.}\ \bibnamefont {Dou}},\ and\ \bibinfo {author} {\bibfnamefont {Q.}~\bibnamefont {Zhao}},\ }\bibfield  {title} {\bibinfo {title} {Quantum battery of interacting spins with environmental noise},\ }\href {https://doi.org/10.1103/PhysRevA.103.033715} {\bibfield  {journal} {\bibinfo  {journal} {Phys. Rev. A}\ }\textbf {\bibinfo {volume} {103}},\ \bibinfo {pages} {033715} (\bibinfo {year} {2021})}\BibitemShut {NoStop}%
\bibitem [{\citenamefont {Grazi}\ \emph {et~al.}(2024)\citenamefont {Grazi}, \citenamefont {Sacco~Shaikh}, \citenamefont {Sassetti}, \citenamefont {Traverso~Ziani},\ and\ \citenamefont {Ferraro}}]{Grazi24}%
  \BibitemOpen
  \bibfield  {author} {\bibinfo {author} {\bibfnamefont {R.}~\bibnamefont {Grazi}}, \bibinfo {author} {\bibfnamefont {D.}~\bibnamefont {Sacco~Shaikh}}, \bibinfo {author} {\bibfnamefont {M.}~\bibnamefont {Sassetti}}, \bibinfo {author} {\bibfnamefont {N.}~\bibnamefont {Traverso~Ziani}},\ and\ \bibinfo {author} {\bibfnamefont {D.}~\bibnamefont {Ferraro}},\ }\bibfield  {title} {\bibinfo {title} {Controlling energy storage crossing quantum phase transitions in an integrable spin quantum battery},\ }\href {https://doi.org/10.1103/PhysRevLett.133.197001} {\bibfield  {journal} {\bibinfo  {journal} {Phys. Rev. Lett.}\ }\textbf {\bibinfo {volume} {133}},\ \bibinfo {pages} {197001} (\bibinfo {year} {2024})}\BibitemShut {NoStop}%
\bibitem [{\citenamefont {Catalano}\ \emph {et~al.}(2024)\citenamefont {Catalano}, \citenamefont {Giampaolo}, \citenamefont {Morsch}, \citenamefont {Giovannetti},\ and\ \citenamefont {Franchini}}]{Catalano24}%
  \BibitemOpen
  \bibfield  {author} {\bibinfo {author} {\bibfnamefont {A.}~\bibnamefont {Catalano}}, \bibinfo {author} {\bibfnamefont {S.}~\bibnamefont {Giampaolo}}, \bibinfo {author} {\bibfnamefont {O.}~\bibnamefont {Morsch}}, \bibinfo {author} {\bibfnamefont {V.}~\bibnamefont {Giovannetti}},\ and\ \bibinfo {author} {\bibfnamefont {F.}~\bibnamefont {Franchini}},\ }\bibfield  {title} {\bibinfo {title} {Frustrating quantum batteries},\ }\href {https://doi.org/10.1103/PRXQuantum.5.030319} {\bibfield  {journal} {\bibinfo  {journal} {PRX Quantum}\ }\textbf {\bibinfo {volume} {5}},\ \bibinfo {pages} {030319} (\bibinfo {year} {2024})}\BibitemShut {NoStop}%
\bibitem [{\citenamefont {Lu}\ \emph {et~al.}(2025)\citenamefont {Lu}, \citenamefont {Tian}, \citenamefont {L\"u},\ and\ \citenamefont {Shang}}]{Lu25}%
  \BibitemOpen
  \bibfield  {author} {\bibinfo {author} {\bibfnamefont {Z.-G.}\ \bibnamefont {Lu}}, \bibinfo {author} {\bibfnamefont {G.}~\bibnamefont {Tian}}, \bibinfo {author} {\bibfnamefont {X.-Y.}\ \bibnamefont {L\"u}},\ and\ \bibinfo {author} {\bibfnamefont {C.}~\bibnamefont {Shang}},\ }\bibfield  {title} {\bibinfo {title} {Topological quantum batteries},\ }\href {https://doi.org/10.1103/PhysRevLett.134.180401} {\bibfield  {journal} {\bibinfo  {journal} {Phys. Rev. Lett.}\ }\textbf {\bibinfo {volume} {134}},\ \bibinfo {pages} {180401} (\bibinfo {year} {2025})}\BibitemShut {NoStop}%
\bibitem [{\citenamefont {Murphy}\ \emph {et~al.}(2025)\citenamefont {Murphy}, \citenamefont {Kiely}, \citenamefont {D'Amico},\ and\ \citenamefont {Campbell}}]{murphy2025ergotopy}%
  \BibitemOpen
  \bibfield  {author} {\bibinfo {author} {\bibfnamefont {D.}~\bibnamefont {Murphy}}, \bibinfo {author} {\bibfnamefont {A.}~\bibnamefont {Kiely}}, \bibinfo {author} {\bibfnamefont {I.}~\bibnamefont {D'Amico}},\ and\ \bibinfo {author} {\bibfnamefont {S.}~\bibnamefont {Campbell}},\ }\bibfield  {title} {\bibinfo {title} {Ergotopy transport in a one dimensional spin chain},\ }\href@noop {} {\bibfield  {journal} {\bibinfo  {journal} {arXiv preprint arXiv:2508.04770}\ } (\bibinfo {year} {2025})}\BibitemShut {NoStop}%
\end{thebibliography}%

\end{document}